\documentclass[AMA,STIX1COL]{WileyNJD-v2}

\articletype{Article Type}%

\received{26 April 2016}
\revised{6 June 2016}
\accepted{6 June 2016}

\newcommand{\beginsupplement}{%
        \setcounter{table}{0}
        \renewcommand{\thetable}{S\arabic{table}}%
        \setcounter{figure}{0}
        \renewcommand{\thefigure}{S\arabic{figure}}%
     }

\begin{document}

\title{A likelihood-based 
sensitivity analysis for publication bias on the
summary ROC in meta-analysis of diagnostic test accuracy}

\author[1]{Yi Zhou}

\author[1]{Ao Huang}

\author[1,2]{Satoshi Hattori*}

\authormark{Zhou \textsc{et al}}

\address[1]{\orgdiv{Department of Biomedical Statistics}, 
\orgname{Graduate School of Medicine, Osaka University}, 
\orgaddress{\state{Osaka}, 
\country{Japan}}}

\address[2]{\orgdiv{Integrated Frontier Research for Medical Science Division}, 
\orgname{Institute for Open and Transdisciplinary Research Initiatives, 
Osaka University}, 
\orgaddress{\state{Osaka}, 
\country{Japan}}}

\corres{Satoshi Hattori, 
Department of Biomedical Statistics, 
Graduate School of Medicine, Osaka University, Osaka, Japan.\\
\email{hattoris@biostat.med.osaka-u.ac.jp}}


\abstract[Summary]{
In meta-analysis of diagnostic test accuracy, 
the summary receiver operating characteristic (SROC) curve is a recommended method to summarize the diagnostic capacity of a medical test in the presence of study-specific cutoff values.
The SROC curve can be estimated by bivariate modeling of pairs of sensitivity and specificity across multiple diagnostic studies, 
and the area under the SROC curve (SAUC) gives the aggregate estimate of diagnostic test accuracy. 
However, publication bias is a major threat to the validity of the estimates. 
To make inference of the impact of publication bias on the SROC curve or the SAUC, we propose a sensitivity analysis method by extending the likelihood-based sensitivity analysis of Copas.
In the proposed method, the SROC curve or the SAUC are estimated by maximizing the likelihood constrained by different values of the marginal probability of selective publication under different mechanisms of selective publication.
A cutoff-dependent selection function is developed to model the selective publication mechanism via the $t$-type statistics or $p$-value of the linear combination of the logit-transformed sensitivity and specificity from the published studies. 
It allows us to model selective publication suggested by the funnel plots of sensitivity, specificity, or diagnostic odds ratio, which are often observed in practice.
A real meta-analysis of diagnostic test accuracy is re-analyzed to illustrate the proposed method, and simulation studies are conducted to evaluate its performance.
}



\maketitle


\section{Introduction}
\label{sec1}

Diagnostic studies play a vital role in evaluating the diagnostic accuracy of medical tests. 
Suppose we are interested in evaluating the diagnostic capacity of a test defined by a continuous variable.
Most diagnostic studies represent the diagnostic capacities by reporting the diagnostic test accuracy measures of sensitivity and specificity pairs, sometimes in combination with the diagnostic odds ratio (DOR). 
These measures are estimated based on the test positive and negative, defined by a cutoff value. 
Although the area under the receiver operating characteristic (ROC) curve (AUC) is also a useful measure of discrimination and is free from the cutoff value, 
the cutoff-dependent sensitivity and specificity pairs are more commonly reported since they are the most intuitive measures of diagnostic accuracy. 
\cite{Ray2010}

The diagnostic studies tend to be conducted in small sample size and present imprecision estimates.
Thus, it is necessary to provide the summaries of data from the relevant studies to get a higher level of evidence on the diagnostic test accuracy.
Meta-analysis of diagnostic test accuracy is an essential statistical tool to synthesize data and estimate the test accuracy from multiple studies.
Each diagnostic study often defines a study-specific cutoff value and then reports sensitivity, specificity, or DOR, which are dependent of the cutoff value.
Thus, it is not appealing to summarize diagnostic capacities by simply averaging these cutoff-dependent quantities over the studies regardless of the variation of cutoff values in the meta-analysis.
\cite{Macaskill2010, Hattori2018}
In the presence of heterogeneous cutoff values, the summary receiver operating characteristic (SROC) curve is a recommended summary measure.
\cite{Hattori2018,Moses1993,Rutter2001,Macaskill2004,Reitsma2005,HaitaoChu2010}
The SROC curve depicts diagnostic test accuracy at all possible cutoff values by presenting the relationship between the sensitivity (true positive rate, TPR) and one minus specificity (false positive rate, FPR) in a monotonic curve.
The area under the SROC curve, the summary AUC (SAUC), gives the average of FPR over all possible values of TPR and is a natural candidate summary.
\cite{Walter2002}
The SROC curve can be estimated by two types of bivariate mixed-effects models. 
One is the bivariate binomial model (hierarchical SROC or HSROC model) for TPR and FPR pairs, relying on the ordinal logistic regression with Bayesian inference or maximum likelihood estimation.
\cite{Rutter2001,Macaskill2004}
The other is the bivariate normal model for logit-transformed sensitivity and specificity pairs, based on the asymptotic bivariate normality of empirical sensitivity and specificity.
\cite{Reitsma2005}

The validity of meta-analysis is threatened by publication bias due to the selective publication of journals.
This phenomenon is widely recognized in meta-analysis of intervention studies, typically randomized clinical trials.
Methods dealing with publication bias have been extensively studied in meta-analysis of intervention studies evaluating the treatment effect.
Graphical methods, such as the funnel plot and the trim-and-fill method, are widely used to detect and adjust for publication bias;
detection is made by evaluating the asymmetry of the funnel plot, and adjustment is made by imputing unpublished studies to make the funnel plot symmetry.
\cite{Macaskill2001}
These graphical methods are simple and provide visual inspection. Then, they are widely used in meta-analysis of intervention studies.
However, the interpretations of these methods can be subjectively misleading.
More quantitative alternative methods have been developed for publication bias by using selection functions to model the selective publication mechanisms.
Copas and colleagues\cite{Copas1999, Copas2000, Copas2001}
successfully introduced methods to model the selective publication mechanisms by applying the Heckman-type\cite{Heckman1976, Heckman1979} selection function, which was first used in econometrics.
In the Heckman-type selection function by Copas and colleagues, 
a latent continuous (Gaussian) random variable was introduced to describe the selective publication.
In intervention studies,
testing hypothesis and corresponding summary statistics like the odds ratio and the hazard ratio between interventions are key quantities in scientific discussion.
Thus, it is a natural way to model the mechanism of selective publication as a function of the corresponding test statistic.
With the Heckman-type selection functions, as seen in equation 4 of Copas and Shi,
\cite{Copas2000}
the probability of publication was a complicated function on the summary statistics and its standard error of the treatment effect.
To quantify the impacts of selective publication in a more intuitively understandable and interpretable matter, 
Copas\cite{Copas2013}
proposed a likelihood-based sensitivity analysis method with a monotonic parametric selection function of the $t$-statistics of the estimated treatment effects.
In medical journals, the $p$-value would be a critical quantity in the scientific arguments and then would be very influential on whether the papers are published or not.
Thus, the $t$-statistic based selection function by Copas\cite{Copas2013}
is an appealing alternative to the Heckman-type selection function.

In contrast, methods for publication bias in meta-analysis of diagnostic test accuracy have been understudied.
Due to the two-dimensional nature of data in diagnostic studies, it is impossible to apply the aforementioned methods directly. 
An intuitive idea is to utilize the graphical methods of the funnel plot and the trim-and-fill method for some univariate diagnostic measures such as the (log-transformed) DOR. 
In many meta-analyses of diagnostic test accuracy, authors assessed publication bias by presenting a funnel plot for the log-transformed DOR (lnDOR).
\cite{Li2013,Zhang2020}
Statistical properties of the funnel-plot-based methods were examined in several articles.
By employing the lnDOR, Deeks et al.\cite{Deeks2005} 
conducted simulation studies to evaluate the usefulness of the Begg, Egger, and Macaskill tests of funnel plot asymmetry.\cite{Macaskill2001,Begg1994,Egger1997}
Their results showed that those tests had low power in detecting publication bias when the cutoff values were heterogeneous,
and they suggested using the regression test of asymmetry with the effective sample size.
B{\"{u}}rkner and Doebler\cite{Burkner2014} 
furthermore compared the performance of the three tests and the trim-and-fill method in combination with four different univariate measures of diagnostic test accuracy.
Their simulation studies showed that the combination of the lnDOR and the trim-and-fill method was the best to detect publication bias, but it lacked power when the number of studies in the meta-analysis was small.
Thus, the graphical methods may be helpful to detect publication bias.
However, they suffer from essential difficulty due to the two-dimensional nature of meta-analysis of diagnostic studies. 
Even if publication bias is detected and suspected, its impact on inference of the SROC curve cannot be addressed. 
Methods based on selection functions are advantageous in quantifying the impact of publication bias on the SROC curve, 
and recently, several developments have been made by using the Heckman-type selection function.
%
Piao et al.\cite{Piao2019} and Li et al.\cite{Li2021} 
proposed the methods based on the conditional and empirical likelihoods, respectively, to correct publication bias based on the bivariate normal model.
Their methods adopted a natural extension of the Heckman-type selection function of Copas and Shi.
\cite{Copas2000, Copas2001}
Hattori and Zhou\cite{Hattori2018}
extended the sensitivity analysis method of Copas and Shi\cite{Copas2000, Copas2001}
to the meta-analysis of diagnostic test accuracy based on the bivariate binomial model.
An explanation of the Heckman-type selection function for the bivariate binomial model was given in Appendix A of Hattori and Zhou.
\cite{Hattori2018} 
The Heckman-type selection functions might model the underlying selective publication process appropriately.
However, 
it cannot explicitly model selective publication process suggested by asymmetry of the funnel plot for the lnDOR.
The funnel plot asymmetry of the lnDOR cannot be explicitly modeled by the Heckman-type selection functions. 
Furthermore, since we cannot completely identify the underlying mechanism of selective publication, 
we should examine the robustness of meta-analysis results against as many mechanisms of selective publication as possible. 
Thus, we need to extend options of selection function.

In this paper, 
we develop an alternative sensitivity analysis method for the meta-analysis of diagnostic test accuracy by utilizing the selection function of $t$-type statistic.
In meta-analysis of diagnostic studies, funnel plot asymmetry is often observed for lnDOR, \cite{Li2013}
and then $t$-statistic for the lnDOR is suggested to be responsible for selective publication.
If this is the case, using selection functions of $t$-type statistic for the lnDOR may be more appealing to model the mechanism of selective publication. 
We introduce a more general class of selection function as a function of the linear combination of the logit-transformed sensitivity and specificity, which includes the lnDOR as a special case. 
We then propose a likelihood-based sensitivity analysis method for the bivariate normal model,
\cite{Reitsma2005}
which is a bivariate extension of the method by Copas.
\cite{Copas2013}
The proposed sensitivity analysis method can quantify the potential bias in the estimation of the SROC curve and the corresponding SAUC by the selective publication driven by the $t$-statistic of lnDOR.
In addition,
by setting suitable coefficients of the linear combination of the logit-transformed sensitivity and specificity, one can model selective publication measured by sensitivity or specificity. 
The proposed method are also applicable with the coefficients unknown.
This flexibility is important since it is very hard to clarify the selective publication mechanisms in diagnostic studies. 
%

The rest of this article is organized as follows. 
In Section \ref{sec2.1}, 
we introduce the notation and definitions of the bivariate normal model without taking into account publication bias, the SROC curve, and the SAUC.
In Section \ref{sec2.2},
we develop the sensitivity analysis method comprising of the selection function and the likelihood function, in the presence of publication bias.
In Section \ref{sec3},
we use a real meta-analysis to illustrate the proposed sensitivity analysis method.
In Section \ref{sec4},
we conducted simulation studies to evaluate the performance of the proposed method 
and to graphically interpret the determinant of publication bias by using the SROC curves.
Finally, we conclude with a discussion in Section \ref{sec5}.
Some results are presented in Supplementary Material.

\section{Bivariate normal model without publication bias}
\label{sec2.1}

Suppose that $N$ diagnostic studies are published and included in meta-analysis.
In this section, 
we assume that all the studies conducted on the test of interest are included, 
or the $N$ studies are random samples from the population of $S(S>N)$ studies.
Each study $i~(i = 1, 2, \dots, N$) reports the observed numbers of true positives, false negatives, true negatives, and false positives, 
denoted by $n_{11}^i, n_{01}^i, n_{00}^i$, and $n_{10}^i$, respectively, as formulated in Table \ref{tab:confusion}.
Let $n_{+1}^i = n_{11}^i + n_{01}^i$ be the number of diseased subjects and $n_{+0}^i =n_{00}^i + n_{10}^i$ the number of non-diseased subjects. 
The observed sensitivity and specificity from each study are estimated by $\mathrm{\hat{se}}_i = {n_{11}^i}/{n_{+1}^i}$ and $\mathrm{\hat{sp}}_i = {n_{00}^i}/{n_{+0}^i}$, respectively.
We use a bivariate normal model\cite{Reitsma2005}
(hereinafter referred to as the Reitsma model) for sensitivity and specificity and define $\mu_{1i}$ and $\mu_{2i}$ as the logit-transformed true sensitivity and specificity of the $i$th study.
%
The Reitsma model assumes that $(\mu_{1i}, \mu_{2i})^T$ is normally distributed:
\begin{align}
	\binom{\mu_{1i}}{\mu_{2i}}
	\sim 
	N\left ( \binom{\mu_1}{\mu_2}, \boldsymbol{\Omega} \right ) 
	\mathrm{~with~}
	\boldsymbol{\Omega} = 
	\begin{pmatrix}
	\tau_1^2 & \tau_{12} \\ 
	\tau_{12} & \tau_2^2
	\end{pmatrix},
	\label{eq:b1}
\end{align}
where $\mu_1$ and $\mu_2$ are the common means of the logit-transformed sensitivity and specificity,
$\tau_1^2~(\tau_1>0)$ and $\tau_2^2~(\tau_2>0)$ are their between-study variances, 
$\tau_{12} = \rho\tau_{1}\tau_{2}$ is the covariance between $\mu_{1i}$ and $\mu_{2i}$,
and $\rho~(-1 \le \rho \le 1)$ is the correlation coefficient.
Let $y_{1i}$ and $y_{2i}$ be the logit-transformed observed sensitivity ($\mathrm{\hat{se}}_i$) and specificity ($\mathrm{\hat{sp}}_i$).
Given $(\mu_{1i}, \mu_{2i})$, it is assumed that
\begin{align}
	\binom{y_{1i}}{y_{2i}} 
	\sim 
	N \left (\binom{\mu_{1i}}{\mu_{2i}}, \boldsymbol{\Sigma}_i  \right )
	\mathrm{~with~}
	\boldsymbol{\Sigma}_i 
	= \begin{pmatrix}
	s_{1i}^2 & 0\\ 
	 0 & s_{2i}^2
	\end{pmatrix},
	\label{eq:b2}
\end{align}
where $s_{1i}^2$ and $s_{2i}^2$ are the observed variances of $y_{1i}$ and $y_{2i}$ within each study.
When $n_{+1}^i$ and $n_{+0}^i$ are large and $0 < \mathrm{\hat{se}}_i;~\mathrm{\hat{sp}}_i < 1$,
the variances can be estimated by $s_{1i}^2 = 1/n_{11}^i + 1/n_{01}^i$ and $s_{2i}^2 = 1/n_{00}^i + 1/n_{10}^i$, respectively.
For studies with frequencies of zero in Table \ref{tab:confusion}, the continuity correction is made by adding 0.5 to all the cells.
Following the convention of literature-based meta-analysis, we regard $\boldsymbol{\Sigma}_i$ as known.
The models (\ref{eq:b1}) and (\ref{eq:b2}) leads to the marginal model:
\begin{align}
\boldsymbol{y}_i | \boldsymbol{\Sigma}_i 
\sim N_2 
\left (\boldsymbol{\mu}, \boldsymbol{\Omega} + \boldsymbol{\Sigma}_i  \right ),
\label{eq:b12}
\end{align}
where $\boldsymbol{y}_i = (y_{1i},y_{2i})^T$, $\boldsymbol{\mu} = (\mu_1,\mu_2)^T$, and $N_2$ denotes the bivariate normal distribution.

The SROC curves derived from the Reitsma model and the HSROC model are statistically rigorous and have been proved to be closely related.
\cite{Harbord2007}
According to the definition of the Reitsma model, the SROC curve can be induced by the conditional expectation of $\mu_{1i}$ given $\mu_{2i}$:
$E(\mu_{1i}|\mu_{2i} = u) = \mu_1 + \dfrac{\tau_{12}}{\tau_2^2}(u - \mu_2)$.
%
Let $x$ be $1 - \mathrm{specificity}$, the SROC curve is defined by
\begin{align}
SROC(x; \boldsymbol{\mu}, \boldsymbol{\Omega}) 
= \mathrm{logit}^{-1} \left[ \mu_1 - \dfrac{\tau_{12}}{\tau_2^2}\{\mathrm{logit}(x)+\mu_2\} \right].
\label{eq:sroc}
\end{align}
Accordingly, the SAUC is defined by 
\begin{align}
SAUC(\boldsymbol{\mu}, \boldsymbol{\Omega}) 
= \int_{0}^{1}SROC(x; \boldsymbol{\mu}, \boldsymbol{\Omega})dx.
\label{eq:sauc}
\end{align}
One can estimate the SROC curve and the SAUC by replacing their theoretical quantities with the maximum likelihood estimators (MLE). 
The HSROC model\cite{Rutter2001}
presents another definition of the SROC curve, hereinafter referred to as the HSROC curve (see equation \ref{eq:hsroc} of appendix for the definition of HSROC). 
With poor estimates of $\rho$ by the Reitsma model, the SROC curve may not be monotone. 
Using correspondence between the Reitsma model and the HSROC,
\cite{Harbord2007}
one may plot the HSROC curve even with the Reitsma model.
\cite{Doebler2020}
The HSROC curve is given by the SROC curve \eqref{eq:sroc} with $\rho=-1$. 
A proof is given in the appendix.
Correspondingly, the area under the HSROC curve (HSAUC) takes the integral of the curve.

\section{Sensitivity analysis for publication bias}
\label{sec2.2}

\subsection{Selection functions on the \textit{t}-type statistic}
\label{sec2.3}

%
To model the selective publication mechanism in diagnostic studies,
we propose a cutoff-dependent selection function alternative to the existing Heckman-type selection functions.
\cite{Hattori2018,Piao2019,Li2021}
For a diagnostic study,
once the cutoff value is fixed, the DOR as a univariate measure of diagnostic capacity combines the strengths of sensitivity and specificity.
\cite{Glas2003}
%
%
The DOR is defined by
\begin{align*}
\mathrm{DOR} = 
\left.\dfrac{\mathrm{se}}{(1-\mathrm{se})}\right/\dfrac{(1-\mathrm{sp})}{\mathrm{sp}},
\end{align*}
and its log-transformation, lnDOR, is
\begin{align*}
\mathrm{lnDOR} = 
\log \dfrac{\mathrm{se} \times \mathrm{sp}}{(1-\mathrm{se}) \times (1-\mathrm{sp})}  
= \mathrm{logit(se)}+\mathrm{logit(sp)}
= y_1 + y_2, 
\end{align*}
which is approximately normally distributed.
The absence of diagnostic capacity corresponds to $\mathrm{DOR}=1$ (equivalently, $\mathrm{lnDOR}=0$).
To test the statistical significance of diagnostic capacity, one can used the $\chi^2$-test for the $2 \times 2$ confusion matrix (Table 1) or the $t$-test for the lnDOR. 
Suppose that all the studies for meta-analysis took a common cutoff value to define the outcomes, 
then selection function on the $t$-statistic of the lnDOR is applicable to model the selective publication mechanism. 
%
%
We consider a more general form of selection function on the $t$-type statistic of the linear combination of the logit-transformed sensitivity and specificity:
\begin{align*}
\boldsymbol{c}^T \boldsymbol{y}_i = c_1y_{1i}+c_2y_{2i},
\end{align*}
where $\boldsymbol{c} = (c_1, c_2)^T$ is a contrast vector.
From equation \eqref{eq:b12}, it holds that
\begin{align}
t_i = \dfrac{\boldsymbol{c}^T \boldsymbol{y}_i}{\sqrt{\boldsymbol{c}^T\boldsymbol{\Sigma}_i\boldsymbol{c}}}
\sim 
N 
\left (
\dfrac{\boldsymbol{c}^T\boldsymbol{\mu}}{\sqrt{\boldsymbol{c}^T\boldsymbol{\Sigma}_i\boldsymbol{c}}},
1 + \dfrac{\boldsymbol{c}^T\boldsymbol{\Omega}\boldsymbol{c}}{\boldsymbol{c}^T\boldsymbol{\Sigma}_i\boldsymbol{c}}
\right ). 
\label{eq:t2} 
\end{align}
Since the $t$-type statistic is scale-invariant, without loss of generality, we constrain that $\boldsymbol{c}^T\boldsymbol{c} = 1~(0 \le c_1; c_2 \le 1)$.
Then, $(c_1, c_2) = (1/\sqrt{2}, 1/\sqrt{2})$ gives the $t$-statistic of the lnDOR.
We then define the selection function $P(\mathrm{select}|\boldsymbol{y}_i, \boldsymbol{\Sigma}_i)$ as a function $a$ of $t_i$:
\begin{align}
P(\mathrm{select}|\boldsymbol{y}_i, \boldsymbol{\Sigma}_i) 
= a(\boldsymbol{y}_i, \boldsymbol{\Sigma}_i)= a(t_i) . 
\label{eq:p}
\end{align}
Copas\cite{Copas2013}
discussed three plausible parametric selection functions, among which the probit function was most tractable.
Since we need to handle more complicated multivariate cases, we also employ the probit function:
\begin{align}
P(\mathrm{select}|\boldsymbol{y}_i, \boldsymbol{\Sigma}_i) 
= a(t_i) = \Phi (\beta t_i + \alpha),
\label{eq:p2}
\end{align}
where $\beta$ and $\alpha$ are the parameters that control the selective publication probability. 
This monotonic parametric model bridges two extreme situations of random selection.
One situation is where random selection with $P(\mathrm{select}) = p_0$ happens when $\beta = 0$ and $\alpha = \Phi^{-1}(p_0)$.
%
%
The other situation is where all the studies are selected with $P(\mathrm{select}) = 1$ when $\beta$ approaches infinity.

By taking different contrast vectors, the $t$-type statistic can determine a variety of selective publication mechanisms.
For example, 
$(c_1, c_2) = (1, 0)$ and $(c_1, c_2) = (0, 1)$ in equation \eqref{eq:t2} indicate that the selective publication mechanisms are determined by the significance of sensitivity and specificity, respectively.

Let the marginal probability of selective publication be $p = P(\mathrm{select})$.
From the selection function \eqref{eq:p}, it holds that:
\begin{align*}
p = P(\mathrm{select}) 
= E_P\{a(t_i)\},
\end{align*}
which indicates the expected proportion of the studies published for meta-analysis from the population.


The definition of the probit function allows us to represent equation \eqref{eq:p2} into
\begin{align}
P(\mathrm{select}|\boldsymbol{y}_i, \boldsymbol{\Sigma}_i) =
a(t_i)
= P(z_i < \beta t_i + \alpha | \boldsymbol{y}_i, \boldsymbol{\Sigma}_i) 
= \Phi \left(
{\beta 
\dfrac{\boldsymbol{c}^T\boldsymbol{y}_i}{\sqrt{\boldsymbol{c}^T\boldsymbol{\Sigma}_i\boldsymbol{c}}}} 
+ \alpha
\right)\label{eq:bp},
\end{align}
where $z_i \sim N(0,1)$ independent of $t_i$.
According to the distribution of $t_i$ in equation \eqref{eq:t2}, the selection function $a(t_i)$ leads to $b(\boldsymbol{\Sigma}_i)$:
\begin{align*}
P(\mathrm{select} | \boldsymbol{\Sigma}_i) 
= b(\boldsymbol{\Sigma}_i)
= P \left (z_i - \beta t_i < \alpha | \boldsymbol{\Sigma}_i \right) 
= \Phi \left\{ \dfrac{ \beta \dfrac{\boldsymbol{c}^T\boldsymbol{\mu}}{ \sqrt{\boldsymbol{c}^T\boldsymbol{\Sigma}_i\boldsymbol{c}}}
+ \alpha}
{  \sqrt{1+\beta^2 \left(1+\dfrac{\boldsymbol{c}^T\boldsymbol{\Omega}\boldsymbol{c}}{\boldsymbol{c}^T\boldsymbol{\Sigma}_i\boldsymbol{c}} \right) }}  \right\}.
\nonumber
\end{align*}

\subsection{Sensitivity analysis}
\label{sec2.4}

Following Copas\cite{Copas2013},
we make inference based on the likelihood function given published, with a fixed marginal probability $p$ as a sensitivity parameter.
We use $f_O$ to denote the distributions defined across $N$ published (observed) studies
and $f_P$ the distributions defined across $S$ population (both published and unpublished) studies.
Given a value of marginal probability of selective publication, $p=P(\mathrm{select})$,
the distribution of the observed $\boldsymbol{\Sigma}_i$ can be obtained by
\begin{align}
f_O(\boldsymbol{\Sigma}_i) 
= P( \boldsymbol{\Sigma}_i | \mathrm{select}) 
= \dfrac{P(\mathrm{select} | \boldsymbol{\Sigma}_i) P(\boldsymbol{\Sigma}_i)}{P(\mathrm{select})} = 
\dfrac{b(\boldsymbol{\Sigma}_i) f_P(\boldsymbol{\Sigma}_i)}{p},
\nonumber
\end{align}
which gives
\begin{align}
f_P(\boldsymbol{\Sigma}_i) = p\dfrac{1}{b(\boldsymbol{\Sigma}_i)}f_O(\boldsymbol{\Sigma}_i).
\label{eq:fps}
\end{align}
When integrating both sides of equation \eqref{eq:fps} over $\boldsymbol{\Sigma}_i$, we can get 
\begin{align}
p = [ E_O\{ b(\boldsymbol{\Sigma}_i)^{-1} \} ]^{-1},
\label{eq:ap}
\end{align}
and it approximately holds that ${p} \approx N/\sum_{i=1}^{N}\{b(\boldsymbol{\Sigma}_i)^{-1}\}$.
Finally, the joint distribution of the observed $(\boldsymbol{y}_i, \boldsymbol{\Sigma}_i)$ can be written as
\begin{align*}
f_O(\boldsymbol{y}_i, \boldsymbol{\Sigma}_i) 
 &= P(\boldsymbol{y}_i, \boldsymbol{\Sigma}_i | \mathrm{select})  \\
 &= \dfrac{f_P(\boldsymbol{y}_i, \boldsymbol{\Sigma}_i)a(\boldsymbol{y}_i, \boldsymbol{\Sigma}_i)}{p}  \\ 
 &= \dfrac{f_P(\boldsymbol{y}_i | \boldsymbol{\Sigma}_i)f_P(\boldsymbol{\Sigma}_i)a(\boldsymbol{y}_i, \boldsymbol{\Sigma}_i)f_O(\boldsymbol{\Sigma}_i)}{b(\boldsymbol{\Sigma}_i)f_P(\boldsymbol{\Sigma}_i)}   \\
 &= \dfrac{f_P(\boldsymbol{y}_i | \boldsymbol{\Sigma}_i)a(\boldsymbol{y}_i, \boldsymbol{\Sigma}_i)f_O(\boldsymbol{\Sigma}_i)}{b(\boldsymbol{\Sigma}_i)}, 
\end{align*}
which gives the log-likelihood based on the published studies: 
\begin{align}
\ell_O(\boldsymbol{\mu},\boldsymbol{\Omega}, \boldsymbol{c}, \beta, \alpha)
& = \log \prod_{i=1}^{N} f_O(\boldsymbol{y}_i, \boldsymbol{\Sigma}_i) \nonumber \\
& = \sum_{i=1}^{N}\log f_P(\boldsymbol{y}_i| \boldsymbol{\Sigma}_i) 
  + \sum_{i=1}^{N}\log a(\boldsymbol{y}_i, \boldsymbol{\Sigma}_i) 
  - \sum_{i=1}^{N}\log b(\boldsymbol{\Sigma}_i) 
  + \sum_{i=1}^{N}\log f_O(\boldsymbol{\Sigma}_i). \label{eq:fllk}
\end{align}
The first term in equation \eqref{eq:fllk} is the log-likelihood of the Reitsma model without taking into account publication bias.
The second and third terms imply the bias correction.
%
The last term is a constant, only dependent on the observed variances, and then does not contribute to maximizing $\ell_O$
To make inference, we fix the marginal selection probability $p$ as a sensitivity. 
Recall that equation \eqref{eq:ap} gives ${p} \approx N/\sum_{s = 1}^{N}\{b(\boldsymbol{\Sigma}_i)^{-1}\}$.
By solving this equation, the parameter $\alpha$ can be written as a function of the remaining parameters 
$(\boldsymbol{\mu},\boldsymbol{\Omega}, \boldsymbol{c}, \beta)$ 
and a given value of $p$.
We denote this by 
$\alpha_p = \alpha_p(\boldsymbol{\mu},\boldsymbol{\Omega}, \boldsymbol{c}, \beta)$.
Finally, the log-likelihood \eqref{eq:fllk} can be represented as a function of the parameters 
$(\boldsymbol{\mu}, \boldsymbol{\Omega}, \boldsymbol{c}, \beta)$, 
given a specified value of the marginal probability of selective publication:
\begin{align}
\ell_O(\boldsymbol{\mu},\boldsymbol{\Omega}, \boldsymbol{c}, \beta)
& = \ell_O(\boldsymbol{\mu},\boldsymbol{\Omega}, \boldsymbol{c}, \beta, \alpha_p) \nonumber \\ 
& = \sum_{i=1}^{N} \left\{ 
-\frac{1}{2} (\boldsymbol{y}_i-\boldsymbol{\mu})^T(\boldsymbol{\Sigma}_i+\boldsymbol{\Omega})^{-1}(\boldsymbol{y}_i-\boldsymbol{\mu})) 
-\frac{1}{2}\log |\boldsymbol{\Sigma}_i+\boldsymbol{\Omega}| 
\right \}  \nonumber \\ 
& + \sum_{i=1}^{N} \log \Phi \left({ \beta 
\dfrac{\boldsymbol{c}^T\boldsymbol{y}_i}{\sqrt{\boldsymbol{c}^T\boldsymbol{\Sigma}_i\boldsymbol{c}}}} 
+ \alpha_p
\right)
- \sum_{i=1}^{N}  \log \Phi \left\{ \dfrac{\beta \dfrac{\boldsymbol{c}^T\boldsymbol{\mu}}{ \sqrt{\boldsymbol{c}^T\boldsymbol{\Sigma}_i\boldsymbol{c}}} +  \alpha_p}
{  \sqrt{1+\beta^2 \left( 1+\dfrac{\boldsymbol{c}^T\boldsymbol{\Omega}\boldsymbol{c}}{\boldsymbol{c}^T\boldsymbol{\Sigma}_i\boldsymbol{c}} \right) }}  \right \}.
\label{eq:llkp}
\end{align}

The parameters can be estimated by maximizing the conditional log-likelihood \eqref{eq:llkp}.
The contrast vector $\boldsymbol{c}$ can be regarded as unknown parameters to be estimated.
The resulting estimators are denoted by $(\hat{\boldsymbol{\mu}}, \hat{\boldsymbol{\Omega}}, \hat{\boldsymbol{c}}, \hat{\beta})$. 
Alternatively, if one is interested in some specific selective publication mechanism,
$\boldsymbol{c}$ can be assigned with a specific value.
The resulting estimators are denoted by $(\hat{\boldsymbol{\mu}}, \hat{\boldsymbol{\Omega}}, \hat{\beta})$.
The asymptotic normality of 
$(\hat{\boldsymbol{\mu}}, \hat{\boldsymbol{\Omega}}, \hat{\boldsymbol{c}}, \hat{\beta})$ or
$(\hat{\boldsymbol{\mu}}, \hat{\boldsymbol{\Omega}}, \hat{\beta})$
follows from the general theory of the maximum likelihood estimation under the assumptions that 
$S$, $n_{+1}^i$, and $n_{+0}^i$ are large.
Their asymptotic variance-covariance matrix can be consistently estimated by the inverse of the empirical Fisher information matrix following the maximum likelihood theory. 
Then, we can construct confidence intervals of the quantities of interest.
For example, a two-tailed confidence interval of either sensitivity or specificity with significance level $\alpha$ can be obtained by
\begin{align*}
\mathrm{logit}^{-1} \left \{
\mu_i \pm  z_{1-\alpha/2} \hat s_{\mu_i}
\right\},
\end{align*}
where $i = 1$ for sensitivity and $i = 2$ for specificity, and $\hat s_{\mu_i}$ denotes the standard error of the logit-transformed sensitivity and specificity, respectively.
We then denote the estimates of the SROC curve and the SAUC as
$SR\hat{O}C = SROC(x; \hat{\boldsymbol{\mu}}, \hat{\boldsymbol{\Omega}})$
and
$SA\hat UC = SAUC(\hat{\boldsymbol{\mu}}, \hat{\boldsymbol{\Omega}})$, respectively.
According to the delta method, the variance of the SAUC is consistently estimated by
\begin{align*}
Var(SA\hat UC) = \hat{\boldsymbol{D}}^T \hat{\boldsymbol{\Sigma}} \hat{\boldsymbol{D}}.
\end{align*}
Here, we define that $\boldsymbol{D} = \int_{0}^{1} SROC(x)\{1-SROC(x)\}\nabla SROC(x) dx$;
$\hat{\boldsymbol{D}}$ denotes the $\boldsymbol{D}$ with its unknown quantities replaced with their MLE,
and $\hat{\boldsymbol{\Sigma}}$ is the estimated variance-covariance matrix of $(\hat{\boldsymbol{\mu}}, \hat{\boldsymbol{\Omega}})$ from the proposed method.
We denote $\nabla SROC(x)$ as the gradient of the SROC curve.
For the SROC curve \eqref{eq:sroc} derived from the Reitsma model,
$\nabla SROC(x)=(1,-\rho\tau_1/\tau_2, -\rho/\tau_2\{\mathrm{logit}(x)+\mu_2\}, \rho\tau_1/\tau_2^2\{\mathrm{logit}(x)+\mu_2\}, -\tau_1/\tau_2\{\mathrm{logit}(x)+\mu_2\})^T$.
%
%
Then, by applying the delta-method to logit-transformed $SA\hat UC$, a two-tailed confidence interval of $SA\hat UC$ can be estimated by
\begin{align*}
\mathrm{logit}^{-1} \left \{
\mathrm{logit} (SA\hat UC) \pm  z_{1-\alpha/2}
\dfrac{\hat{s}_{SAUC}}{SA\hat UC(1-SA\hat UC)} \right \},
\end{align*}
where $\hat{s}_{SAUC} = \sqrt{Var(SA\hat UC)}$.
The logit-transformation restricts that the confidence interval of $SA\hat UC$ is in the interval $[0,1]$.
Recall that the above inference was based on a fixed $p$. 
Since no one know the true marginal probability of selective publication $p$, it is recommended to examine how sensitive the $SA\hat UC$ is with a range of $p$'s.

\section{Application}
\label{sec3}

\begin{figure}[!htb]
\centering\includegraphics[width=0.8\columnwidth]{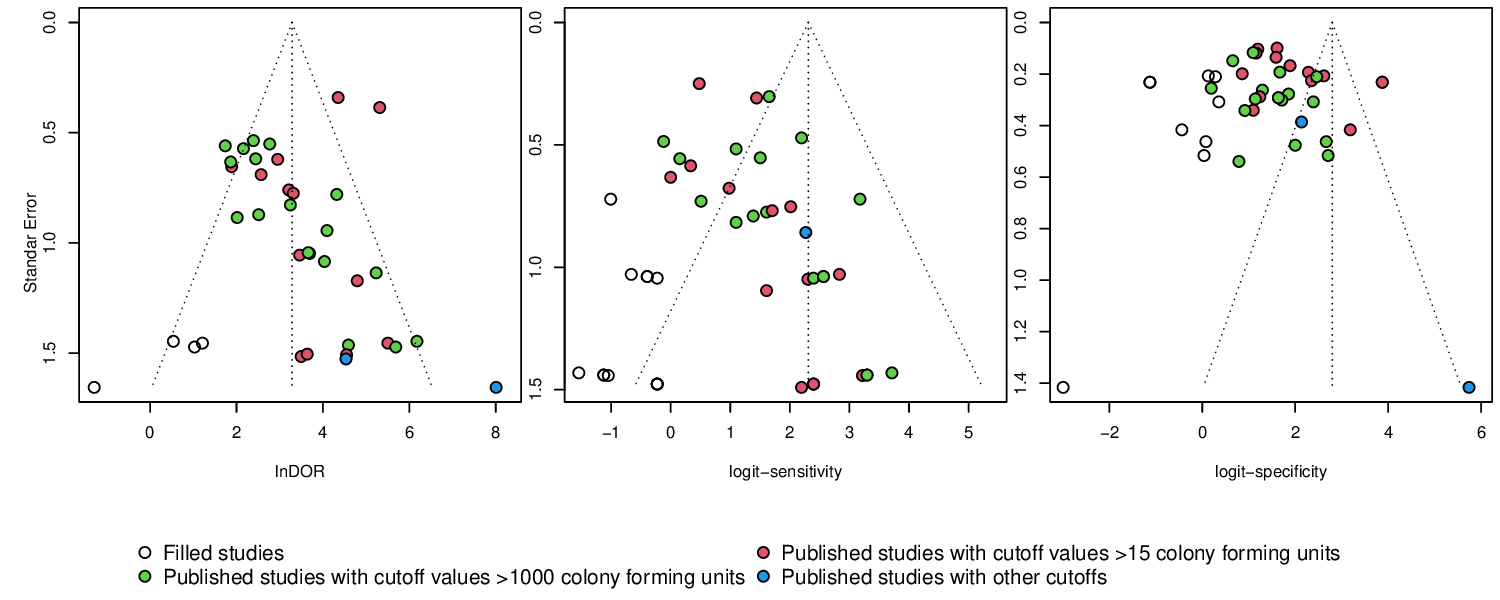}
\caption{
The trim-and-fill results on the lnDOR, logit-transformed sensitivity, and specificity in IVD example.
The dashed lines are the estimates of the trim-and-fill method.}
\label{fig:IVD-FUNNEL}
\end{figure}

We illustrate the proposed sensitivity analysis method by reanalyzing the published meta-analysis used in Hattori and Zhou.
\cite{Hattori2018}
This meta-analysis investigated 33 diagnostic studies of semi-quantitative and quantitative catheter segment culture tests for assessing the test accuracy in diagnosing intravascular device (IVD) related bloodstream infection,
and the test positives in the studies were defined by different criteria.
\cite{Safdar2005}
In Figure \ref{fig:IVD-FUNNEL}, funnel plots of the lnDOR, logit-transformed sensitivity, and logit-transformed specificity against their standard errors were presented.
All the plots in Figure \ref{fig:IVD-FUNNEL} showed asymmetry, 
suggesting that some selective publication might exist, and all of the lnDOR, logit-transformed sensitivity, and logit-transformed specificity have the potential to describe it. 
We applied the proposed method to estimate the SROC curves derived from the Reitsma model given the marginal probability of selective publication $p = 1, 0.8, 0.6, 0.4$.
%
Recall that the marginal probability of selective publication indicates the expected proportion of the published studies in the population.
When $p=1$, there are no unpublished studies, 
and the inference of the proposed method by the maximum likelihood estimation agrees with that of the Reitsma model without taking into account publication bias.
When $p=0.4$, for example, the expected number of the unpublished studies is $(1-0.4) \times 33/0.4 \approx 50$, in the presence of moderate publication bias.

\begin{figure}[!hbt]
\centering\includegraphics[width=0.8\columnwidth]{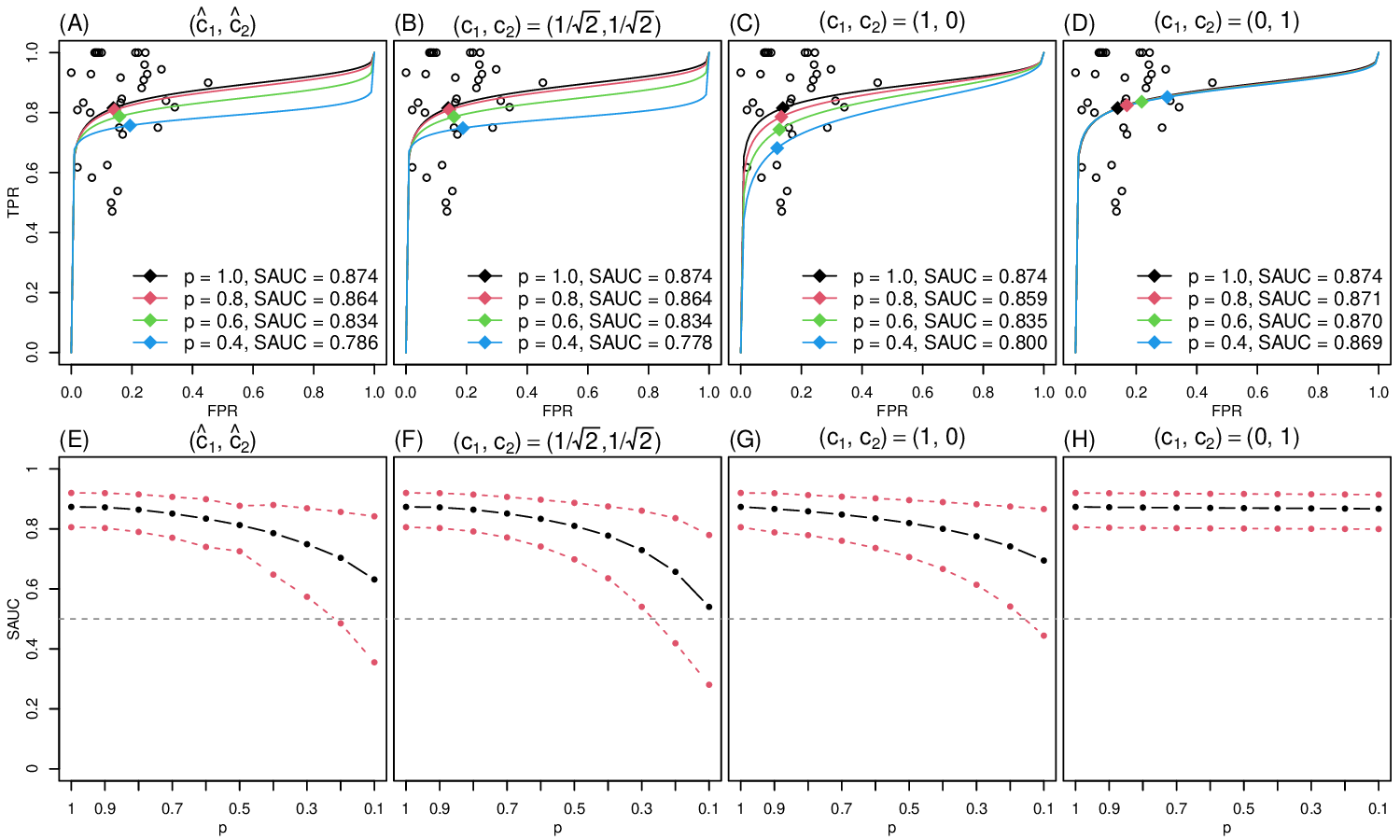}
\caption{
The estimated SROC curves under four scenarios of selective publication mechanism in IVD example.
The circle points are the observed FPR and TPR pairs from 33 primary studies.
The diamond points are the estimated SOP.
In panel (A), 
$(\hat{c}_1, \hat{c}_2) = (0.746, 0.666), (0.691, 0.723), (0.657, 0.754)$ given $p = 0.8,0.6,0.4$, respectively.
For all the $p$'s, $\hat{c}_1\approx \hat{c}_2$ suggests lnDOR is responsible for selective publication.
}
\label{fig:IVD-SROC}
\end{figure}

Since it was hard to clarify what kind of selective publication mechanism existed from the funnel plots in Figure \ref{fig:IVD-FUNNEL},
we first did not assume any specific selective publication mechanism in this meta-analysis, or regarded $(c_1, c_2)$ as unknown.
The results of the estimated SROC curves under the mechanisms of $(\hat c_1, \hat c_2)$ are shown in panel (A) of Figure \ref{fig:IVD-SROC}.
As $p$ decreases from 1 to 0.4, the corresponding SAUC estimates decreased from 0.874 to 0.786.
Since we could not conclude that the mechanism of $(\hat c_1, \hat c_2)$ was correct,  
we need to consider some other mechanisms to see the robustness of the results and conjecture the determinant of publication bias.
We then specified another three selective publication mechanisms:
$(c_1, c_2) = (1/\sqrt{2}, 1/\sqrt{2}), (1,0)$, and $(0,1)$.
These three mechanisms investigated how the estimated SROC curves and the SAUC would change over $p$'s if publication bias was assumed to be determined by the significance of the DOR, sensitivity, and specificity, respectively.
Their corresponding estimated SROC curves are presented in panels (B)-(D) of Figure \ref{fig:IVD-SROC}.
As given in the caption of Figure \ref{fig:IVD-SROC}, the estimated values of $(\hat{c}_1, \hat{c}_2)$ showed $\hat c_1\approx \hat c_2$ for all the $p$'s,
and then the mechanism of $(c_1, c_2)=(1/\sqrt{2}, 1/\sqrt{2})$ was supported.
Correspondingly, results with $(\hat{c}_1, \hat{c}_2)$ were similar to those with $c_1= c_2$ 
%
%
When publication bias was assumed to be determined by the $t$-type statistic on sensitivity (i.e., $(c_1, c_2) = (1, 0)$) or specificity (i.e., $(c_1, c_2) = (0, 1)$), 
the results of the SROC curves became different, which means that different considerations of $(c_1, c_2)$ in the selection function certainly influenced the estimates.
Thus, to investigate the robustness of results, we should take into account different selective publication mechanisms.

In Figure \ref{fig:IVD-SROC},
we presented the summary operating points (SOPs), which were $(\hat\mu_1, \hat\mu_2)$ for each $p$.
Under the mechanisms of $(\hat{c}_1, \hat{c}_2)$ and $(c_1, c_2)=(1/\sqrt{2},1/\sqrt{2})$, the SOPs changed similarly on the SROC space, as shown in panels (A) and (B) of Figure \ref{fig:IVD-SROC}.
The results suggested that, under the selection function with $c_1 = c_2$, studies with low sensitivity and low specificity located in the right lower part of the SROC space were likely to be unpublished.
Under the assumed mechanism of $(c_1, c_2)=(1,0)$, 
the estimated FPR of the SOPs did not change much over $p$'s, whereas the estimated TPR decreased, as shown in panel (C) of Figure \ref{fig:IVD-SROC}.
The opposite results were obtained under $(c_1, c_2)=(0,1)$, as shown in panel (D) of Figure \ref{fig:IVD-SROC}.
From these results, one may expect that tracing the SOPs would help understand the selective publication mechanism determined by the specified selection function.
The estimated TPR and FPR of the SOPs and their 95\% confidence intervals are presented in Table S1 of the Supplementary Material.

\begin{figure}[!hbt]
\centering\includegraphics[width=0.8\columnwidth]{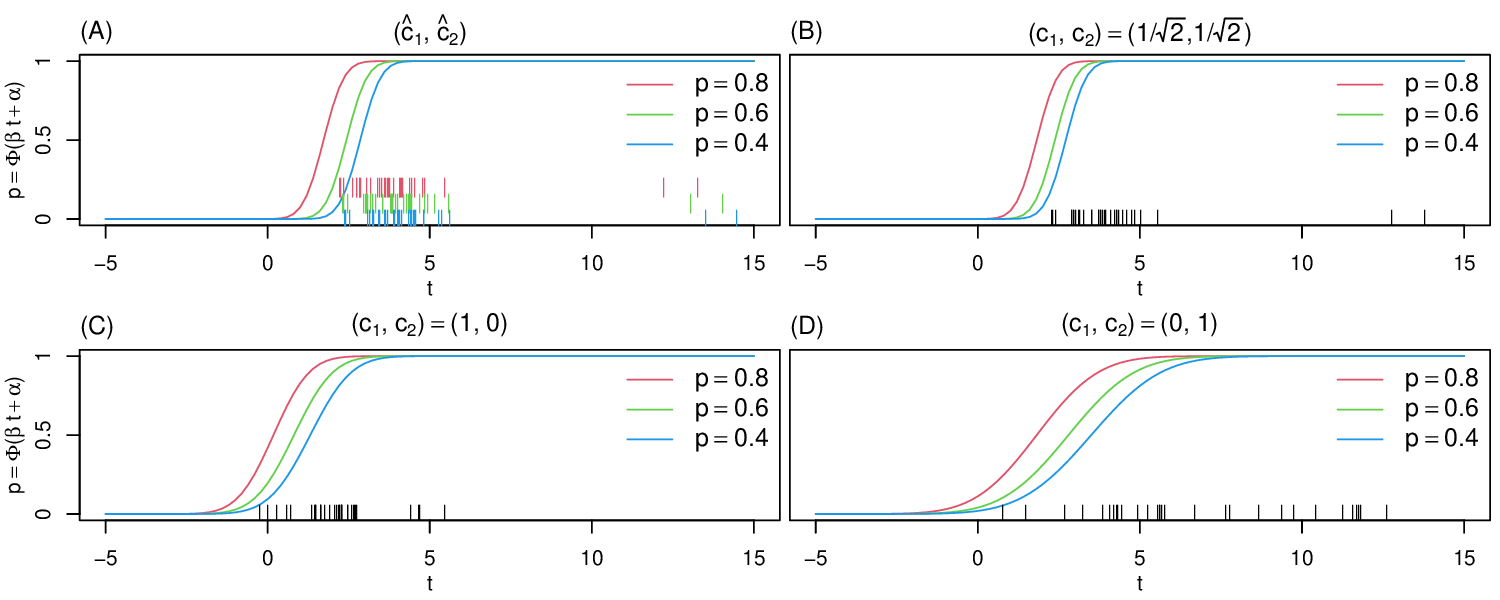}
\caption{
The estimated probit selection function $a(t_i)$ given different $p$'s in IVD example.
The vertical lines at the bottom are the estimated $t$-type statistics from 33 published studies.
}
\label{fig:IVD-PROBIT}
\end{figure}

To examine the influence of $(\hat c_1, \hat c_2)$ on the precision of the results,
we presented the SAUC estimates with the 95\% confidence intervals under the aforementioned mechanisms over $p=1, 0.9, \dots, 0.1$ in panel (E)-(H) of Figure \ref{fig:IVD-SROC}
The confidence intervals obtained under $(\hat c_1, \hat c_2)$ were slightly wider than those under $(c_1, c_2)=(1/\sqrt{2},1/\sqrt{2})$, indicating that estimating $(c_1, c_2)$ seemed not to lead serious loss in precision.
With regard to the proposed selection function, we plotted the estimated probit selection functions on the $t$-statistics of 33 published studies in Figure \ref{fig:IVD-PROBIT}.
%
Under the mechanisms of $(\hat{c}_1, \hat{c}_2)$ and $(c_1, c_2)=(1/\sqrt{2},1/\sqrt{2})$, almost all the published studies had high selection probability, 
indicating that the estimated selection functions were reasonable to describe the probability of selective publication of the studies.
When the $t$-statistics were only dependent on sensitivity or specificity, in panels (C) or (D) of Figure \ref{fig:IVD-PROBIT} respectively,
the shape of the estimated selection functions became much flatter.

As given in Figure \ref{fig:IVD-FUNNEL}, 
the funnel plot was not helpful to identify which measure was most suitable to describe the mechanism of selective publication
or to make quantitative evaluations.
In contrast, the proposed method showed the advantages of 
the ability to quantify the possible impacts of publication bias on the SROC curve and the SAUC,
and give more insights on the mechanism of the selective publication by tracing the trajectories of SOPs.
In summary, under four selective publication mechanisms and a range of $p$'s, 
the estimated test accuracy measure of the SAUC maintained statistically significant, 
even if the marginal probability of selective publication was not high like $p=0.3$ or 77 unpublished studies.
The results of the proposed sensitive analysis strengthened the conclusion that semi-quantitative and quantitative catheter segment culture tests had high sensitivity and specificity to diagnose IVD related bloodstream infections.

An additional application illustrating the proposed sensitivity analysis method on the HSROC curve or the HSAUC was presented in the Supplementary Material.

\section{Simulation studies}
\label{sec4}

\subsection{Performance of estimation on the SAUC}
\label{sec4.1}

Simulation studies were conducted to assess the performance of the estimation by the proposed sensitivity analysis method. 
%
We considered small size $(S=15, 25)$, moderate size $(S=50)$, and large size $(S=200)$ cases of the meta-analysis.
Recall that $S$ is the number of all the studies (published and unpublished).
The large sample size might be impractical but was considered to check the performance of the proposed method in an ideal situation.
True sensitivity and specificity were set as $(0.5, 0.85)$, $(0.8, 0.8)$, or $(0.9, 0.4)$, yielding $(\mu_1, \mu_2)$ to be $(0, 1.735)$, $(1.386, 1.386)$, or $(2.197, -0.405)$, respectively.
The between-study variances $(\tau_1^2, \tau_2^2)$ were set as $(1, 4)$, and their correlation $\rho$ was set as $-0.3$ or $-0.6$.
The within-study variances $s_{1}^2$ or $s_{2}^2$ were randomly generated from the square of $N(0.5, 0.5^2)$.
\cite{Piao2019}
We took into account the true selective publication mechanisms with the contrast vector $(c_1, c_2)$ to be $(1/\sqrt{2}, 1/\sqrt{2})$, $(1, 0)$, or $(0, 1)$
which define the $t$-type statistics on the lnDOR, sensitivity, or specificity, respectively.
In the true selection function $a(t_i)$ (equation \ref{eq:bp}), we set $\beta$ to be 0.5, and $\alpha$ was calculated to ensure that the marginal probability of selective publication approximately equal to 0.7, which means about 70\% of studies were published from $S$ studies.
The mixture of these settings produced six scenarios for the population data in each of the three selective publication mechanisms.  
The scenarios were summarized in Table \ref{tab:sce12}.

%
In each scenario, we generated 1000 sets of published studies as follows:
\begin{enumerate}[Step 1:]
\item generate $S$ population studies from $\boldsymbol{y}_i \sim N_2 \left( \boldsymbol{\mu}, \boldsymbol{\Omega} + \boldsymbol{\Sigma}_i \right)$,
with $s_{1i}$ or $s_{2i} \sim N(0.5, 0.5^2)$, $i = 1, 2, \dots, S $;
\item calculate the probability of each study $i$ being selected in the meta-analysis,
$p_i = \Phi (\beta {t}_i + \alpha)$,
and generate the random indicator $z_i \sim \mathrm{Bernoulli}(p_i)$;
\item select the studies with indicator $z_i = 1$ as one set of published studies. 
\end{enumerate}

We estimated the SAUC by using the MLE of Reitsma model \eqref{eq:b12} with $N$ published studies, as well as with $S$ population studies. 
The latter is used as an ideal reference.
We compared our estimates with them.
To apply the proposed method, the marginal probability $p$ was set as $p =1/S\sum_{s = 1}^{S}p_i$. 
This is not practically applicable, but we evaluated the performances of our methods with the suitably specified $p$.
We assessed the SAUC estimates when the contrast vector $(c_1, c_2)$ was estimated, correctly specified by the true values, and misspecified.
The misspecification of $(c_1, c_2)$ was used to verify that misspecifying the contrast vector would cause somewhat biased estimates.
We also compared with the method by Piao et al.,\cite{Piao2019} 
which used the Heckman-type selection function.
All statistical computing was conducted by R (R Development Core Team, Version 4.0.5).
The Reitsma model was conducted by R package mixmeta (version 1.1.3).
\cite{Sera2019}
The proposed conditional likelihood was numerically optimized by using R function nlminb().
One potential concern in the optimization is that the initial values or the constrained bounds for $\beta$ may have some impact on the validity of estimates.
To check these issues, we compared the results from several reasonable initial values and reasonable constrained bounds for $\beta$, and the results did not differ much.

The median, the 25th empirical percentile, and the 75th empirical percentile of the estimated SAUC were summarized.
The estimates by different methods under the true selective publication mechanism of 
$(c_1,c_2)=(1/\sqrt{2},1/\sqrt{2})$, $(1, 0)$, and $(0, 1)$ are presented in Table \ref{tab:c11}, Table \ref{tab:c10}, and Table \ref{tab:c01}, respectively.
In the true selective publication mechanism of $(c_1, c_2) = (0, 1)$, only specificity was the determinant of publication bias,
bringing about the random selection on sensitivity.
Since the SAUC measures the average of sensitivity given specificity, the impact of publication bias on the SAUC was not obvious in the true mechanism of 
$(c_1, c_2) = (0, 1)$.
In contrast, publication bias on the SAUC was significant in the mechanisms of $(c_1, c_2) = (1/\sqrt{2}, 1/\sqrt{2})$ and $(c_1, c_2) = (1,0)$, 
especially in the former mechanism that both sensitivity and specificity gave rise to severe publication bias.
The overall results of the estimated SAUC showed that the proposed methods under $(\hat c_1, \hat c_2)$ and the correctly specified $(c_1, c_2)$ considerably removed the publication bias from the Reitsma model based on the published studies.
Less than 5\% bias could be observed when the DOR or only sensitivity was the determinant of the publication bias (Table \ref{tab:c11}-\ref{tab:c01}).
When only specificity was the determinant of the publication bias, 
the proposed method with $(c_1, c_2)$ correctly specified had small bias (Table \ref{tab:c01}), 
while the proposed method with $(\hat{c}_1, \hat{c}_2)$ had some moderate bias (less than 10\%) when the true SAUC was low and the population size was small (Scenario No. 1 and No. 2 of Table \ref{tab:c01}).
The moderate bias mainly resulted from some biased estimates of $(\hat{c}_1, \hat{c}_2)$, 
which consequently caused the bias in the parameters,
when the size of the population studies was not large enough ($S=15, 25, 50$).
In the true selective publication mechanism of $(c_1, c_2) = (1/\sqrt{2}, 1/\sqrt{2})$ and $(0, 1)$,
the proposed methods with $(\hat{c}_1, \hat{c}_2)$ or correct specification were comparable and even outperformed the Heckman-type method,
because the Heckman-type method possibly suffered from the misspecification under the specified mechanism.
When $(c_1, c_2)$ was misspecified in the proposed method, biased estimates were presented in all the scenarios.
We also calculated the convergence rates, which is the success rate of iterations in maximizing the log-likelihood \eqref{eq:llkp} to obtain the proposed estimates. 
The proposed methods with $(\hat{c}_1, \hat{c}_2)$ or correct specification could obtain convergence rates close to 100\% under all the scenarios.
%
%

Since the magnitude of $\tau_1^2$ or $\tau_2^2$ might be related with the bias on sensitivity or specificity and might influence the magnitude of publication bias on the SAUC, 
we additionally considered the scenarios of small between-study variances $(\tau_1^2, \tau_2^2) = (0.5, 0.5)$ in simulation studies.
The scenarios are summarized in Table S3 of the Supplementary Material.
When $\tau_1^2$ or $\tau_2^2$ were small, 
the publication bias on the SAUC decreased in general, 
and the summaries of the SAUC estimates under the three kinds of selective publication mechanisms were in agreement with those of $(\tau_1^2, \tau_2^2) = (1, 4)$, 
as shown in Table S4-S6 of the Supplementary Material.

\subsection{Tracing the SOP to speculate the underlying selective publication mechanism}
\label{sec4.2}

\begin{figure}[!hbt]
\centering\includegraphics[width=0.8\columnwidth]{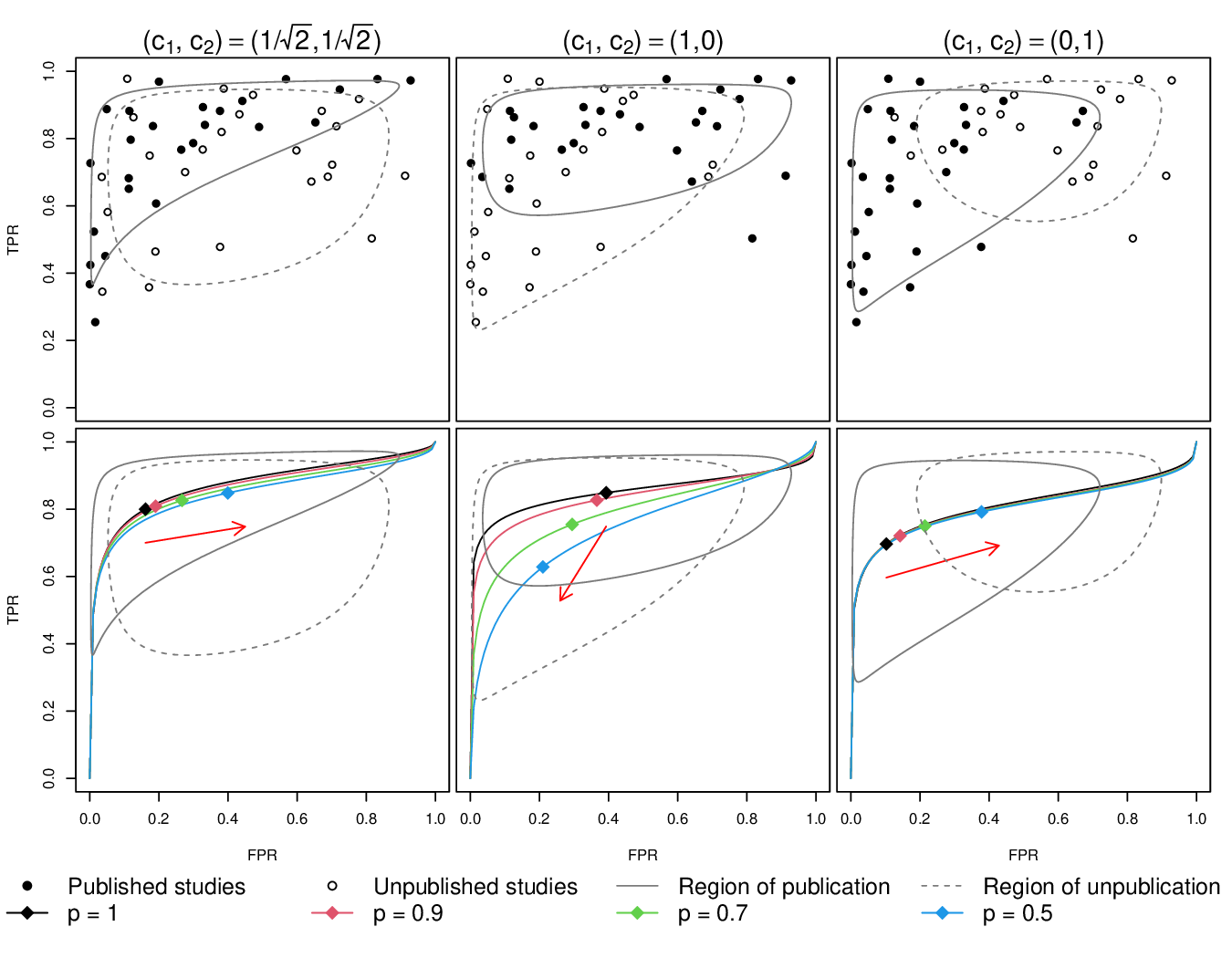}
\caption{
Changes of the estimated SROC curves and the SOP when $(c_1, c_2)$ are correctly specified.
The published or unpublished studies are simulated from Scenario No. 3.
The diamond points are the estimated SOP.
The regions of publication and unpublication were given by the logistic-transformed ellipse,
of which the centers are the average of logit-transformed FPR and TPR pairs from the primary studies,
and the shapes are based on their covariance matrix.
}
\label{fig:sim-sauc}
\end{figure}

As suggested in Section \ref{sec3}, tracking the trajectory of the SOP over the different marginal probability of selective publication $p$'s might give us some insights into the underlying selective publication mechanism.
We randomly picked one dataset of meta-analysis with $S=50$ under Scenario No. 3 in Table \ref{tab:sce12} to illustrate how the SOP estimates could track the unpublished studies.
We considered the selective publication mechanism of $(c_1, c_2) =(1/\sqrt{2}, 1/\sqrt{2})$, $(1, 0)$, and $(0, 1)$ with $p = 0.5$.
Based on the published studies, the SROC curves and the SOP were estimated using the proposed method that correctly specifies $(c_1, c_2)$, given $p=1, 0.9, 0.7, 0.5$.
In the upper panels of Figure \ref{fig:sim-sauc}, the scatter plots of the published (filled circle) and unpublished (open circle) studies are shown in the SROC space. 
The region, where published studies are, is shown with a solid curve and that for unpublished studies is with a broken curve.
In the lower panels, the estimated SROC curves over $p=1, 0.9, 0.7, 0.5$ and the trajectories of the SOP are shown.
%
All the trajectories of SOP seemed to move from the regions of the published studies to those of the unpublished studies,
indicating that tracing the trajectories of the SOP is useful to identify the region of unpublished studies in the SROC space.

\section{Discussion}
\label{sec5}

Among the recent literature about meta-analysis of diagnostic test accuracy, 
the test of funnel plot asymmetry proposed by Deeks et al.\cite{Deeks2005} is often used to detect publication bias.
However, Deeks test has low power in detecting publication bias when the cutoff values are heterogeneous, and it cannot provide any information on the impact of publication bias on the results.
\cite{Deeks2005,Burkner2014}
Up to date, limited statistical investigation of publication bias has been recommended.
\cite{Salameh2020}
Thus, there is an urgent demand for the method dealing with publication bias in meta-analysis of diagnostic test accuracy.

In meta-analysis of intervention studies, 
sensitivity analysis methods based on selection functions provide more objective and stable ways to quantify and adjust publication bias than the widely used funnel plot and the trim-and-fill method.
\cite{Carpenter2009,Schwarzer2010}
Recently, sensitivity analysis methods have been developed for publication bias in more complicated situations, including network meta-analysis
\cite{Mavridis2013}
and diagnostic studies.
\cite{Hattori2018,Piao2019,Li2021}
%
Since graphical methods such as the funnel plot and the trim-and-fill method are not appealing for meta-analysis of diagnostics studies due to the two-dimensional nature of data, 
the importance of sensitivity analysis with selection functions should be more emphasized. 
Several sensitivity analysis methods have been proposed so far, and they are all based on the Heckman-type selection functions.
In this article, we took a different approach by extending the likelihood-based sensitivity analysis of Copas.
\cite{Copas2013}
In the proposed sensitivity analysis method, 
we allow the selection function to depend on the study-specific cutoff values, 
and thus the developed selection function can model different selective publication mechanisms from the Heckman-type.
Consequently, the proposed method gives statistical investigations on the impact of publication bias on the SROC curves or the SAUC based on various selective publication processes.
This feature would be appealing for diagnostic studies because the primary outcomes are usually reported depending on the study-specific cutoff values, 
and the cutoff values inevitably cause variety of the selective publication process.

Even with these advantages, we must mention several limitations.
%
The proposed method is based on the Reitsma model,
which uses asymptotic bivariate normality of the empirical sensitivity and specificity.
An ad hoc continuity correction is needed when zero frequencies are found in the number of true positives, true negatives, false positives, or false negatives in a study.
%
When the number of studies or the number of the diseased or the non-diseased subjects is very small, 
the bivariate binomial model is a more natural model than the Reitsma model to incorporate the within-study variability and does not need the ad hoc continuity correction.
For the bivariate binomial model, only one Heckman-type selection model has been applied for sensitivity analysis.
\cite{Hattori2018}
It would be interesting and important to extend our method to the bivariate binomial model.
The proposed sensitivity analysis method allows incorporating $t$-type statistic of the linear combination of logit-transformed sensitivity and specificity to represent the different selective publication mechanisms,
for example, the mechanisms determined by sensitivity, specificity, or the DOR.
However, it is not easy to verify the underlying selective publication mechanism in the meta-analysis, and it is also unclear whether a common measure like DOR can be the determinant of publication bias for all the studies.
The intuitive idea is to allow choosing different types of statistics for each study and define a selection function through its $p$-value, 
but it would be hard to decide which statistic is responsible for publication for each study.
Accumulating practical experience and clinical interpretation would be warranted.
Since it may be impossible to identify the underlying selective publication mechanism, 
one approach is to conduct several sensitivity analyses under the assumptions of different selective publication mechanisms by assigning different values to the contrast vector.
As demonstrated in Section \ref{sec3} and \ref{sec4.2}, 
tracing the SOPs might be useful to make interpretations of the assumed publication mechanism by the specified selection function. 
On the other hand, it is helpful to estimate the contrast vector to reduce the uncertainty on the underlying selection process.
The simulation studies also revealed that the contrast vector could be stably estimated in the proposed method.
However, even if the contrast vector could be estimated successfully, the estimated results should be considered as the suggestion of the most possible selective publication mechanism rather than the definite estimation.
Since the estimates of the contrast vector are all based on the published studies, if the unpublished studies employed different cutoff values, the locations of the unpublished studies on SROC space can be considerably changed.
For these reasons, it is still important to evaluate potential biases with several specified contrast vectors as a part of sensitivity analysis to draw robust conclusions for the meta-analysis.


\section*{Acknowledgments}

The authors would like to thank the editor, associate editor, and reviewers for their helpful and insightful comments.
This research was partly supported by Grant-in-Aid for Challenging Exploratory Research (16K12403) and for Scientific Research (16H06299, 18H03208) from the Ministry of Education, Science, Sports and Technology of Japan.





\section*{Data available statement}
R codes 
together with a sample
input data set and complete documentation is available at
\url{https://github.com/meta2020/dtametasa-r}.
The proposed methods have been implemented in an R package can be installed from
\url{https://meta2020.github.io/dtametasa/}.

\section*{Supporting information}

Supplementary Material is available as part of the online article.

\bibliography{refs}%
\clearpage


\begin{center}
\begin{table}[!hbt]
\caption{Notations for cell frequencies in $2 \times 2$ confusion matrix  
for diagnostic study $i$.
\label{tab:confusion}}
\centering
\begin{tabular*}{0.5\linewidth}{@{\extracolsep\fill}rrrr@{\extracolsep\fill}}
\toprule
\multicolumn{2}{l}{\multirow{2}{*}{}} & \multicolumn{2}{c}{Truth} \\ \cline{3-4} 
\multicolumn{2}{l}{}                  & Diseased  	   & Non-diseased  \\ 
\midrule
Diagnostic test        & Positive     & $n_{11}^i$ & $n_{10}^i$\\
\multirow{2}{*}{}      & Negative     & $n_{01}^i$ & $n_{00}^i$\\
\addlinespace
                       & Total        & $n_{+1}^i$ & $n_{+0}^i$\\ 
\bottomrule                       
\end{tabular*}
\end{table}
\end{center}
\begin{table}[!hbt]
\caption{\label{tab:sce12}Scenarios for simulation studies given $p \approx 0.7$}
\centering
\begin{tabular}[t]{rrrrrrrrrrrr}
\toprule
\multicolumn{1}{c}{ } & \multicolumn{1}{c}{ } & \multicolumn{1}{c}{ } & \multicolumn{1}{c}{ } & \multicolumn{1}{c}{ } & \multicolumn{1}{c}{ } & \multicolumn{1}{c}{ } & \multicolumn{1}{c}{ } & \multicolumn{1}{c}{ } & \multicolumn{1}{c}{$c_1 = c_2$} & \multicolumn{1}{c}{$c_1 = 1$} & \multicolumn{1}{c}{$c_1 = 0$} \\
\cmidrule(l{3pt}r{3pt}){10-10} \cmidrule(l{3pt}r{3pt}){11-11} \cmidrule(l{3pt}r{3pt}){12-12}
No. & SAUC & $\mu_1$ & $\mu_2$ & $\tau_1^2$ & $\tau_2^2$ & $\tau_{12}$ & $\rho$ & $\beta$ & $\alpha_{0.7}$ & $\alpha_{0.7}$ & $\alpha_{0.7}$\\
\midrule
1 & 0.564 & 0.000 & 1.735 & 1 & 4 & -0.6 & -0.3 & 0.5 & -0.165 & 0.891 & -0.429\\
2 & 0.620 & 0.000 & 1.735 & 1 & 4 & -1.2 & -0.6 & 0.5 & -0.251 & 0.894 & -0.433\\
3 & 0.828 & 1.386 & 1.386 & 1 & 4 & -0.6 & -0.3 & 0.5 & -0.766 & -0.570 & -0.111\\
4 & 0.846 & 1.386 & 1.386 & 1 & 4 & -1.2 & -0.6 & 0.5 & -0.848 & -0.573 & -0.118\\
5 & 0.892 & 2.197 & -0.405 & 1 & 4 & -0.6 & -0.3 & 0.5 & -0.198 & -1.269 & 1.744\\
6 & 0.877 & 2.197 & -0.405 & 1 & 4 & -1.2 & -0.6 & 0.5 & -0.284 & -1.269 & 1.733\\
\bottomrule
\end{tabular}
\end{table}

\begin{table}[!hbt]
\caption{\label{tab:c11}Summary of the SAUC estimates under the true selective publication mechanism of $(c_1, c_2) = (1/\sqrt{2}, 1/\sqrt{2})$}
\centering
\begin{tabular}[t]{rrrrrrr}
\toprule
\multicolumn{1}{c}{} & \multicolumn{1}{c}{} & \multicolumn{1}{c}{} & \multicolumn{1}{c}{$S = 15$} & \multicolumn{1}{c}{$S = 25$} & \multicolumn{1}{c}{$S = 50$} & \multicolumn{1}{c}{$S = 200$} \\
\cmidrule(l{3pt}r{3pt}){4-4} \cmidrule(l{3pt}r{3pt}){5-5} \cmidrule(l{3pt}r{3pt}){6-6} \cmidrule(l{3pt}r{3pt}){7-7}
No. &  Methods & True & Median (Q1, Q3) & Median (Q1, Q3) & Median (Q1, Q3) & Median (Q1, Q3)\\
\midrule
1 & Proposed $(\hat{c}_1, \hat{c}_2)$ & 56.4 & 60.7 (47.4, 70.6) & 59.1 (49.6, 67.1) & 57.2 (49.2, 63.7) & 55.7 (51.5, 59.8)\\
 & Proposed $(c_1 = c_2)$ &  & 59.9 (48.4, 69.6) & 59.4 (51.0, 67.2) & 58.0 (51.9, 64.0) & 56.5 (53.0, 59.6)\\
 & Proposed $(c_1 = 1)$ &  & 65.7 (55.6, 73.5) & 65.9 (59.1, 71.8) & 66.2 (61.5, 70.1) & 66.3 (64.2, 68.6)\\
 & Heckman-type &  & 65.1 (56.0, 72.8) & 65.5 (58.3, 71.3) & 64.9 (59.8, 69.1) & 64.5 (61.9, 66.6)\\
 & Reitsma$_O$ &  & 67.4 (57.8, 74.6) & 67.7 (61.8, 73.0) & 67.2 (63.0, 71.1) & 66.9 (64.8, 69.0)\\
 & Reitsma$_P$ &  & 56.6 (49.5, 63.4) & 56.4 (51.3, 60.7) & 56.3 (52.9, 59.6) & 56.3 (54.6, 58.1)\\
\addlinespace
2 & Proposed $(\hat{c}_1, \hat{c}_2)$ & 62.0 & 64.6 (54.7, 72.7) & 62.8 (54.8, 69.8) & 62.0 (55.7, 67.2) & 61.5 (58.0, 64.7)\\
 & Proposed $(c_1 = c_2)$ &  & 64.5 (55.4, 72.3) & 63.3 (56.0, 69.9) & 63.1 (58.1, 68.2) & 62.1 (59.5, 64.8)\\
 & Proposed $(c_1 = 1)$ &  & 69.0 (61.2, 75.0) & 68.8 (64.1, 73.3) & 69.5 (65.5, 72.1) & 69.4 (67.7, 70.8)\\
 & Heckman-type &  & 67.7 (59.7, 74.0) & 67.4 (62.1, 71.8) & 67.3 (63.4, 71.0) & 67.2 (65.2, 69.0)\\
 & Reitsma$_O$ &  & 69.9 (62.9, 75.7) & 69.5 (65.2, 73.5) & 70.0 (66.5, 72.7) & 69.7 (68.1, 71.2)\\
 & Reitsma$_P$ &  & 62.5 (56.9, 67.3) & 61.9 (57.4, 65.5) & 62.1 (59.2, 64.6) & 62.0 (60.7, 63.2)\\
\addlinespace
3 & Proposed $(\hat{c}_1, \hat{c}_2)$ & 82.8 & 85.0 (79.2, 88.6) & 84.0 (78.9, 87.3) & 83.6 (79.9, 86.2) & 83.0 (81.3, 84.6)\\
 & Proposed $(c_1 = c_2)$ &  & 84.4 (78.7, 88.1) & 83.6 (78.7, 87.0) & 83.5 (80.3, 86.0) & 82.9 (81.3, 84.5)\\
 & Proposed $(c_1 = 1)$ &  & 85.7 (80.4, 89.0) & 85.6 (81.5, 88.2) & 85.8 (83.4, 87.7) & 86.3 (85.1, 87.4)\\
 & Heckman-type &  & 86.1 (82.0, 89.2) & 86.0 (82.9, 88.3) & 86.6 (84.4, 88.1) & 86.4 (85.5, 87.3)\\
 & Reitsma$_O$ &  & 87.0 (83.3, 89.8) & 86.8 (83.9, 89.2) & 87.2 (85.3, 88.7) & 87.2 (86.4, 88.0)\\
 & Reitsma$_P$ &  & 82.6 (78.5, 85.5) & 82.6 (79.9, 85.0) & 82.8 (81.0, 84.4) & 82.8 (81.9, 83.7)\\
\addlinespace
4 & Proposed $(\hat{c}_1, \hat{c}_2)$ & 84.6 & 85.8 (82.2, 88.6) & 85.3 (82.4, 87.7) & 85.0 (82.6, 86.9) & 84.7 (83.6, 85.8)\\
 & Proposed $(c_1 = c_2)$ &  & 85.4 (81.7, 88.6) & 85.2 (82.3, 87.5) & 85.1 (82.9, 87.0) & 84.7 (83.7, 85.8)\\
 & Proposed $(c_1 = 1)$ &  & 86.4 (83.6, 89.2) & 86.4 (84.0, 88.4) & 86.8 (84.9, 88.1) & 87.0 (86.3, 87.8)\\
 & Heckman-type &  & 86.5 (83.6, 89.0) & 86.6 (84.5, 88.4) & 86.7 (85.2, 88.1) & 86.7 (85.9, 87.4)\\
 & Reitsma$_O$ &  & 87.3 (84.6, 89.8) & 87.4 (85.5, 89.1) & 87.6 (86.2, 88.8) & 87.6 (86.9, 88.1)\\
 & Reitsma$_P$ &  & 84.5 (82.0, 86.8) & 84.5 (82.4, 86.5) & 84.7 (83.2, 85.8) & 84.6 (84.0, 85.2)\\
\addlinespace
5 & Proposed $(\hat{c}_1, \hat{c}_2)$ & 89.2 & 88.8 (85.9, 91.4) & 89.3 (87.2, 91.2) & 89.2 (87.6, 90.6) & 89.2 (88.4, 89.9)\\
 & Proposed $(c_1 = c_2)$ &  & 88.8 (86.1, 91.3) & 89.5 (87.4, 91.3) & 89.3 (87.8, 90.6) & 89.2 (88.5, 89.9)\\
 & Proposed $(c_1 = 1)$ &  & 88.4 (85.1, 91.2) & 88.8 (86.2, 91.1) & 89.3 (87.6, 90.8) & 89.8 (88.7, 90.6)\\
 & Heckman-type &  & 89.3 (86.5, 91.6) & 89.8 (87.8, 91.6) & 89.8 (88.4, 91.0) & 89.9 (89.3, 90.6)\\
 & Reitsma$_O$ &  & 90.0 (87.6, 92.1) & 90.6 (88.9, 92.1) & 90.7 (89.6, 91.7) & 90.8 (90.3, 91.3)\\
 & Reitsma$_P$ &  & 88.8 (86.4, 90.9) & 89.1 (87.4, 90.7) & 89.1 (87.9, 90.2) & 89.1 (88.6, 89.7)\\
\addlinespace
6 & Proposed $(\hat{c}_1, \hat{c}_2)$ & 87.7 & 87.5 (84.4, 90.1) & 87.7 (85.4, 89.5) & 87.7 (86.1, 89.1) & 87.8 (87.0, 88.5)\\
 & Proposed $(c_1 = c_2)$ &  & 87.6 (84.9, 90.2) & 88.0 (85.9, 89.7) & 88.0 (86.5, 89.3) & 87.8 (87.1, 88.5)\\
 & Proposed $(c_1 = 1)$ &  & 87.1 (83.8, 89.9) & 87.7 (85.2, 89.4) & 87.7 (85.9, 89.2) & 88.1 (87.1, 88.9)\\
 & Heckman-type &  & 87.7 (85.0, 90.0) & 88.0 (86.0, 89.7) & 87.9 (86.7, 89.2) & 88.0 (87.3, 88.7)\\
 & Reitsma$_O$ &  & 88.5 (85.8, 90.8) & 88.9 (87.3, 90.5) & 89.0 (87.9, 90.1) & 89.1 (88.5, 89.6)\\
 & Reitsma$_P$ &  & 87.5 (85.0, 89.7) & 87.6 (85.9, 89.3) & 87.7 (86.5, 88.8) & 87.7 (87.1, 88.3)\\
\bottomrule
\end{tabular}
\begin{tablenotes}
\item 
Median with 25th empirical percentile (Q1) and 75th empirical percentile (Q3) are reported. 
No. corresponds to the scenario number.
$S$ denotes the number of the population studies. 
True denotes the the true value of the SAUC.
Proposed $(\hat c_1, \hat c_2)$, Proposed $(c_1 = c_2)$, and Proposed $(c_1 = 1)$ denote 
the proposed method that estimates $(c_1, c_2)$, 
correctly specifies $(c_1, c_2) = (1/\sqrt{2}, 1/\sqrt{2})$, 
and misspecifies $(c_1, c_2) = (1,0)$, respectively;
Heckman-type denotes the method of Piao et al.;
Reitsma$_O$ and Reitsma$_P$ denote the Reitsma model 
based on $N$ published studies and $S$ population studies, respectively. 
All the entries are multiplied by 100.
\end{tablenotes}
\end{table}

\begin{table}[!hbt]
\caption{\label{tab:c10}Summary of the SAUC estimates under the true selective publication mechanism of $(c_1, c_2) = (1,0)$}
\centering
\begin{tabular}[t]{rrrrrrr}
\toprule
\multicolumn{1}{c}{} & \multicolumn{1}{c}{} & \multicolumn{1}{c}{} & \multicolumn{1}{c}{$S = 15$} & \multicolumn{1}{c}{$S = 25$} & \multicolumn{1}{c}{$S = 50$} & \multicolumn{1}{c}{$S = 200$} \\
\cmidrule(l{3pt}r{3pt}){4-4} \cmidrule(l{3pt}r{3pt}){5-5} \cmidrule(l{3pt}r{3pt}){6-6} \cmidrule(l{3pt}r{3pt}){7-7}
No. &   & True & Median (Q1, Q3) & Median (Q1, Q3) & Median (Q1, Q3) & Median (Q1, Q3)\\
\midrule
1 & Proposed $(\hat{c}_1, \hat{c}_2)$ & 56.4 & 60.2 (51.7, 67.4) & 59.9 (52.0, 65.7) & 56.8 (51.3, 63.0) & 56.2 (53.3, 58.6)\\
 & Proposed $(c_1 = 1)$ &  & 59.8 (51.3, 67.8) & 59.0 (51.9, 65.3) & 57.3 (52.1, 62.3) & 56.5 (53.9, 58.9)\\
 & Proposed $(c_1 = c_2)$ &  & 62.0 (55.0, 68.4) & 62.7 (57.1, 67.1) & 62.7 (59.1, 66.3) & 64.0 (62.1, 65.8)\\
 & Heckman-type &  & 61.5 (55.1, 68.3) & 62.1 (56.4, 66.6) & 61.8 (58.1, 65.1) & 61.5 (59.5, 63.2)\\
 & Reitsma$_O$ &  & 64.7 (58.8, 70.8) & 65.2 (60.1, 69.3) & 65.2 (61.7, 68.2) & 65.0 (63.3, 66.4)\\
 & Reitsma$_P$ &  & 56.6 (49.5, 63.4) & 56.4 (51.3, 60.7) & 56.3 (52.9, 59.6) & 56.3 (54.6, 58.1)\\
\addlinespace
2 & Proposed $(\hat{c}_1, \hat{c}_2)$ & 62.0 & 64.0 (56.8, 69.9) & 63.0 (57.4, 68.0) & 62.2 (57.6, 66.2) & 62.0 (59.8, 64.0)\\
 & Proposed $(c_1 = 1)$ &  & 64.2 (56.9, 70.0) & 63.1 (57.9, 67.9) & 62.8 (58.6, 66.1) & 62.1 (60.1, 63.9)\\
 & Proposed $(c_1 = c_2)$ &  & 65.0 (58.7, 70.3) & 64.9 (60.5, 68.8) & 65.5 (62.2, 68.1) & 66.7 (65.0, 68.1)\\
 & Heckman-type &  & 64.7 (58.8, 69.7) & 64.0 (59.3, 68.0) & 64.2 (61.1, 67.2) & 64.0 (62.3, 65.6)\\
 & Reitsma$_O$ &  & 67.4 (62.1, 72.4) & 67.3 (63.4, 70.7) & 67.4 (64.7, 69.8) & 67.4 (66.1, 68.7)\\
 & Reitsma$_P$ &  & 62.5 (56.9, 67.3) & 61.9 (57.4, 65.5) & 62.1 (59.2, 64.6) & 62.0 (60.7, 63.2)\\
\addlinespace
3 & Proposed $(\hat{c}_1, \hat{c}_2)$ & 82.8 & 83.9 (79.2, 87.4) & 83.6 (80.0, 86.6) & 83.2 (80.7, 85.4) & 83.0 (81.5, 84.6)\\
 & Proposed $(c_1 = 1)$ &  & 83.4 (78.6, 87.1) & 83.4 (79.1, 86.4) & 83.2 (80.8, 85.5) & 82.9 (81.6, 84.3)\\
 & Proposed $(c_1 = c_2)$ &  & 84.3 (80.2, 87.6) & 84.5 (81.6, 87.0) & 84.8 (82.8, 86.4) & 85.3 (84.2, 86.2)\\
 & Heckman-type &  & 85.3 (81.4, 87.9) & 85.4 (82.8, 87.6) & 85.7 (84.0, 87.2) & 85.8 (85.0, 86.6)\\
 & Reitsma$_O$ &  & 85.9 (82.2, 88.6) & 86.0 (83.7, 88.1) & 86.3 (84.7, 87.6) & 86.4 (85.7, 87.1)\\
 & Reitsma$_P$ &  & 82.6 (78.5, 85.5) & 82.6 (79.9, 85.0) & 82.8 (81.0, 84.4) & 82.8 (81.9, 83.7)\\
\addlinespace
4 & Proposed $(\hat{c}_1, \hat{c}_2)$ & 84.6 & 85.5 (82.4, 88.1) & 85.1 (82.5, 87.3) & 84.9 (83.1, 86.5) & 84.9 (83.9, 85.8)\\
 & Proposed $(c_1 = 1)$ &  & 85.2 (82.0, 87.9) & 85.0 (82.3, 87.2) & 84.9 (83.2, 86.5) & 84.7 (83.8, 85.6)\\
 & Proposed $(c_1 = c_2)$ &  & 85.7 (82.6, 88.2) & 85.6 (83.3, 87.6) & 85.8 (84.2, 87.1) & 85.9 (85.1, 86.7)\\
 & Heckman-type &  & 85.9 (83.3, 88.3) & 86.0 (84.0, 87.8) & 86.3 (84.9, 87.5) & 86.3 (85.6, 86.9)\\
 & Reitsma$_O$ &  & 87.0 (84.2, 89.0) & 86.9 (84.9, 88.5) & 87.0 (85.8, 88.1) & 87.1 (86.5, 87.6)\\
 & Reitsma$_P$ &  & 84.5 (82.0, 86.8) & 84.5 (82.4, 86.5) & 84.7 (83.2, 85.8) & 84.6 (84.0, 85.2)\\
\addlinespace
5 & Proposed $(\hat{c}_1, \hat{c}_2)$ & 89.2 & 90.1 (87.4, 92.2) & 90.3 (88.1, 91.9) & 89.8 (88.4, 91.1) & 89.4 (88.5, 90.2)\\
 & Proposed $(c_1 = 1)$ &  & 89.5 (86.2, 91.9) & 89.6 (87.1, 91.5) & 89.4 (87.6, 90.8) & 89.2 (88.3, 90.0)\\
 & Proposed $(c_1 = c_2)$ &  & 90.7 (88.2, 92.6) & 91.0 (89.3, 92.4) & 91.0 (89.9, 92.0) & 91.1 (90.7, 91.7)\\
 & Heckman-type &  & 91.0 (88.4, 92.9) & 91.2 (89.6, 92.6) & 91.3 (90.2, 92.3) & 91.4 (90.8, 91.8)\\
 & Reitsma$_O$ &  & 91.1 (88.5, 92.9) & 91.5 (89.9, 92.7) & 91.5 (90.4, 92.4) & 91.5 (91.0, 91.9)\\
 & Reitsma$_P$ &  & 88.8 (86.4, 90.9) & 89.1 (87.4, 90.7) & 89.1 (87.9, 90.2) & 89.1 (88.6, 89.7)\\
\addlinespace
6 & Proposed $(\hat{c}_1, \hat{c}_2)$ & 87.7 & 88.7 (86.0, 91.0) & 89.1 (87.0, 90.8) & 88.7 (87.3, 90.0) & 88.3 (87.4, 89.1)\\
 & Proposed $(c_1 = 1)$ &  & 88.0 (84.6, 90.5) & 88.3 (85.8, 90.3) & 88.0 (86.2, 89.4) & 87.8 (87.0, 88.5)\\
 & Proposed $(c_1 = c_2)$ &  & 89.1 (86.8, 91.4) & 89.6 (87.8, 91.2) & 89.5 (88.4, 90.7) & 89.6 (89.0, 90.2)\\
 & Heckman-type &  & 89.3 (86.7, 91.4) & 89.6 (87.9, 91.2) & 89.6 (88.4, 90.7) & 89.7 (89.1, 90.3)\\
 & Reitsma$_O$ &  & 89.5 (87.1, 91.6) & 90.0 (88.2, 91.4) & 89.9 (88.8, 91.0) & 90.0 (89.4, 90.5)\\
 & Reitsma$_P$ &  & 87.5 (85.0, 89.7) & 87.6 (85.9, 89.3) & 87.7 (86.5, 88.8) & 87.7 (87.1, 88.3)\\
\bottomrule
\end{tabular}
\begin{tablenotes}
\item 
Median with 25th empirical percentile (Q1) and 75th empirical percentile (Q3) are reported. 
No. corresponds to the scenario number.
$S$ denotes the number of population studies.
True denotes the true value of the SAUC.
Proposed $(\hat c_1, \hat c_2)$, Proposed $(c_1 = 1)$, and Proposed $(c_1 = c_2)$ denote 
the proposed method that estimates $(c_1, c_2)$, 
correctly specifies $(c_1, c_2)=(1,0)$, 
and misspecifies $(c_1, c_2)=(1/\sqrt{2}, 1/\sqrt{2})$, respectively;
Heckman-type denotes the method of Piao et al.;
Reitsma$_O$ and Reitsma$_P$ denote the Reitsma model 
based on $N$ published studies and $S$ population studies, respectively. 
All the entries are multiplied by 100.
\end{tablenotes}
\end{table}

\begin{table}[!hbt]
\caption{\label{tab:c01}Summary of the SAUC estimates under the true selective publication mechanism of $(c_1, c_2) = (0,1)$}
\centering
\begin{tabular}[t]{rrrrrrr}
\toprule
\multicolumn{1}{c}{} & \multicolumn{1}{c}{} & \multicolumn{1}{c}{} & \multicolumn{1}{c}{$S = 15$} & \multicolumn{1}{c}{$S = 25$} & \multicolumn{1}{c}{$S = 50$} & \multicolumn{1}{c}{$S = 200$} \\
\cmidrule(l{3pt}r{3pt}){4-4} \cmidrule(l{3pt}r{3pt}){5-5} \cmidrule(l{3pt}r{3pt}){6-6} \cmidrule(l{3pt}r{3pt}){7-7}
No. &   & True & Median (Q1, Q3) & Median (Q1, Q3) & Median (Q1, Q3) & Median (Q1, Q3)\\
\midrule
1 & Proposed $(\hat{c}_1, \hat{c}_2)$ & 56.4 & 48.7 (34.5, 62.1) & 48.8 (37.3, 59.9) & 48.0 (38.6, 56.7) & 52.4 (47.6, 56.5)\\
 & Proposed $(c_1 = 0)$ &  & 56.6 (44.2, 67.4) & 56.9 (47.5, 65.2) & 56.5 (50.5, 62.0) & 56.0 (52.9, 59.1)\\
 & Proposed $(c_1 = c_2)$ &  & 47.9 (33.7, 61.4) & 47.8 (37.8, 58.9) & 46.6 (38.6, 55.4) & 47.4 (41.8, 53.8)\\
 & Heckman-type &  & 54.1 (41.2, 65.7) & 54.1 (43.6, 62.8) & 53.9 (47.1, 60.2) & 53.5 (49.8, 57.4)\\
 & Reitsma$_O$ &  & 57.0 (44.3, 68.5) & 57.5 (47.6, 65.9) & 57.1 (50.7, 62.8) & 56.7 (53.5, 59.9)\\
 & Reitsma$_P$ &  & 56.6 (49.5, 63.4) & 56.4 (51.3, 60.7) & 56.3 (52.9, 59.6) & 56.3 (54.6, 58.1)\\
\addlinespace
2 & Proposed $(\hat{c}_1, \hat{c}_2)$ & 62.0 & 57.6 (44.2, 69.1) & 55.6 (43.8, 64.8) & 55.8 (48.0, 62.7) & 59.0 (54.6, 62.5)\\
 & Proposed $(c_1 = 0)$ &  & 63.4 (52.7, 71.8) & 62.3 (54.5, 68.8) & 62.5 (57.4, 66.8) & 62.0 (59.5, 64.1)\\
 & Proposed $(c_1 = c_2)$ &  & 56.2 (43.5, 68.1) & 54.6 (44.1, 64.3) & 55.6 (47.4, 62.5) & 55.7 (50.3, 60.5)\\
 & Heckman-type &  & 61.0 (49.1, 70.2) & 59.9 (51.1, 67.1) & 60.0 (53.6, 65.5) & 59.6 (56.2, 62.5)\\
 & Reitsma$_O$ &  & 64.3 (53.5, 72.6) & 63.3 (55.2, 69.7) & 63.5 (58.1, 68.0) & 63.1 (60.7, 65.3)\\
 & Reitsma$_P$ &  & 62.5 (56.9, 67.3) & 61.9 (57.4, 65.5) & 62.1 (59.2, 64.6) & 62.0 (60.7, 63.2)\\
\addlinespace
3 & Proposed $(\hat{c}_1, \hat{c}_2)$ & 82.8 & 80.0 (67.5, 86.2) & 79.5 (71.8, 85.2) & 79.9 (74.0, 84.2) & 81.9 (79.5, 83.7)\\
 & Proposed $(c_1 = 0)$ &  & 82.6 (74.4, 87.5) & 82.6 (77.5, 86.6) & 82.8 (79.0, 85.6) & 82.8 (81.3, 84.2)\\
 & Proposed $(c_1 = c_2)$ &  & 78.4 (66.3, 85.7) & 78.6 (70.3, 84.6) & 78.3 (72.8, 83.2) & 79.7 (76.3, 82.4)\\
 & Heckman-type &  & 81.0 (72.1, 86.9) & 80.8 (74.8, 85.6) & 81.2 (76.8, 84.7) & 81.4 (79.3, 83.0)\\
 & Reitsma$_O$ &  & 82.9 (73.9, 87.8) & 82.9 (77.5, 87.0) & 83.2 (79.1, 85.9) & 83.2 (81.6, 84.7)\\
 & Reitsma$_P$ &  & 82.6 (78.5, 85.5) & 82.6 (79.9, 85.0) & 82.8 (81.0, 84.4) & 82.8 (81.9, 83.7)\\
\addlinespace
4 & Proposed $(\hat{c}_1, \hat{c}_2)$ & 84.6 & 83.5 (76.3, 87.6) & 82.9 (77.8, 86.4) & 83.2 (79.5, 85.8) & 84.1 (82.7, 85.3)\\
 & Proposed $(c_1 = 0)$ &  & 85.0 (79.9, 88.3) & 84.5 (80.9, 87.2) & 84.6 (82.4, 86.7) & 84.6 (83.6, 85.6)\\
 & Proposed $(c_1 = c_2)$ &  & 83.1 (75.9, 87.4) & 82.0 (76.2, 86.1) & 83.1 (78.9, 85.7) & 83.4 (81.4, 85.1)\\
 & Heckman-type &  & 83.6 (77.2, 87.4) & 83.1 (78.5, 86.5) & 83.3 (80.0, 85.6) & 83.3 (81.9, 84.5)\\
 & Reitsma$_O$ &  & 85.5 (80.1, 88.7) & 85.0 (81.5, 87.7) & 85.1 (82.9, 87.2) & 85.2 (84.1, 86.2)\\
 & Reitsma$_P$ &  & 84.5 (82.0, 86.8) & 84.5 (82.4, 86.5) & 84.7 (83.2, 85.8) & 84.6 (84.0, 85.2)\\
\addlinespace
5 & Proposed $(\hat{c}_1, \hat{c}_2)$ & 89.2 & 87.0 (82.3, 89.9) & 87.8 (84.8, 90.0) & 88.2 (86.3, 89.7) & 88.9 (88.1, 89.6)\\
 & Proposed $(c_1 = 0)$ &  & 88.3 (85.2, 90.7) & 88.9 (86.6, 90.7) & 89.0 (87.6, 90.2) & 89.1 (88.5, 89.7)\\
 & Proposed $(c_1 = c_2)$ &  & 86.9 (82.6, 89.8) & 87.6 (84.8, 89.8) & 87.9 (86.0, 89.5) & 88.6 (87.5, 89.3)\\
 & Heckman-type &  & 87.3 (83.9, 90.1) & 87.7 (85.2, 90.0) & 87.8 (86.2, 89.3) & 87.9 (87.1, 88.7)\\
 & Reitsma$_O$ &  & 88.2 (85.0, 90.7) & 89.0 (86.8, 90.8) & 89.2 (87.8, 90.4) & 89.3 (88.7, 89.9)\\
 & Reitsma$_P$ &  & 88.8 (86.4, 90.9) & 89.1 (87.4, 90.7) & 89.1 (87.9, 90.2) & 89.1 (88.6, 89.7)\\
\addlinespace
6 & Proposed $(\hat{c}_1, \hat{c}_2)$ & 87.7 & 86.3 (82.4, 89.0) & 86.5 (84.2, 88.6) & 87.2 (85.3, 88.5) & 87.5 (86.7, 88.2)\\
 & Proposed $(c_1 = 0)$ &  & 87.1 (84.3, 89.6) & 87.5 (85.5, 89.3) & 87.7 (86.3, 88.9) & 87.7 (87.0, 88.3)\\
 & Proposed $(c_1 = c_2)$ &  & 86.1 (82.7, 89.0) & 86.6 (84.1, 88.7) & 87.0 (85.2, 88.4) & 87.5 (86.7, 88.2)\\
 & Heckman-type &  & 85.8 (82.8, 88.6) & 86.2 (83.7, 88.2) & 86.2 (84.6, 87.7) & 86.2 (85.4, 87.0)\\
 & Reitsma$_O$ &  & 87.2 (84.4, 89.6) & 87.7 (85.7, 89.4) & 87.9 (86.6, 89.1) & 88.0 (87.4, 88.5)\\
 & Reitsma$_P$ &  & 87.5 (85.0, 89.7) & 87.6 (85.9, 89.3) & 87.7 (86.5, 88.8) & 87.7 (87.1, 88.3)\\
\bottomrule
\end{tabular}
\begin{tablenotes}
\item 
Median with 25th empirical percentile (Q1) and 75th empirical percentile (Q3) are reported. 
No. corresponds to the scenario    number.
$S$ denotes the number of the population studies. 
True denotes the the true value of the SAUC.
Proposed $(\hat c_1, \hat c_2)$, Proposed $(c_1 = 1)$, and Proposed $(c_1 = c_2)$ denote 
the proposed method that estimates $(c_1, c_2)$, 
correctly specifies $(c_1, c_2) = (0,1)$, 
and misspecifies $(c_1, c_2) = (1/\sqrt{2}, 1/\sqrt{2})$, respectively;
Heckman-type denotes the method of Piao et al.;
Reitsma$_O$ and Reitsma$_P$ denote the Reitsma model 
based on $N$ published studies and $S$ population studies, respectively. 
All the entries are multiplied by 100.
\end{tablenotes}
\end{table}

\clearpage
\begin{appendix}

\section{The HSROC curve derived from the HSROC model\label{app1}}

In this section, we give a proof on the correspondence between the SROC by the Reitsma model\cite{Reitsma2005} and the HSROC model\cite{Rutter2001}, 
discussed at the end of Section \ref{sec2.1} of the main text.
The unification of the SROC curves derived from these two models
has been discussed by Harbord and colleagues.\cite{Harbord2007}
The SROC curve from the HSROC model, hereinafter, the HSROC curve, is defined by allowing the threshold parameter $\theta_i$ to vary 
while holding the accuracy parameter $\alpha_i$ fixed at its mean $\Lambda$. 
\cite{Rutter2001}
%
For the HSROC model without covariates, equation (5.1) in Harbord et al.\cite{Harbord2007} 
gives expected sensitivity given specificity,
which induces the definition of HSROC.
Let $x$ be $1 - \mathrm{specificity}$,
the HSROC is defined by
\begin{align}
HSROC(x) 
= \mathrm{logit}^{-1} \left[ \Lambda e^{-\beta/2} + e^{-\beta} \mathrm{logit} (x)\} \right].
\label{eq:hsroc}
\end{align}

Under the assumption that $\theta_i$ and $\alpha_i$ are uncorrelated,
Harbord and colleagues derived that $\beta = \log(\tau_2/\tau_1)$
and $\Lambda = \sqrt{\tau_2/\tau_1}\mu_1 + \sqrt{\tau_1/\tau_2}\mu_2$,
where $\tau_1, \tau_2, \mu_1$, and $\mu_2$ are the parameters in the Reitsma model.
After substituting for $\beta$ and $\Lambda$ in equation \ref{eq:hsroc},
we can get 
\begin{align*}
HSROC(x) 
& = \mathrm{logit}^{-1} \left[ \left (\sqrt{\dfrac{\tau_2}{\tau_1}}\mu_1 + \sqrt{\dfrac{\tau_1}{\tau_2}}\mu_2 \right) 
\exp \left\{-\dfrac{1}{2}\log\left(\dfrac{\tau_2}{\tau_1}\right) \right\}
+ \exp \left\{-\log\left(\dfrac{\tau_2}{\tau_1}\right) \right\}
 \mathrm{logit} (x)\right]\\
& = \mathrm{logit}^{-1} \left\{ \left (\sqrt{\dfrac{\tau_2}{\tau_1}}\mu_1 + \sqrt{\dfrac{\tau_1}{\tau_2}}\mu_2 \right) 
\sqrt{\dfrac{\tau_1}{\tau_2}}
+ \dfrac{\tau_1}{\tau_2}
 \mathrm{logit} (x) \right\}\\
 & = \mathrm{logit}^{-1} \left[ \mu_1 + \dfrac{\tau_1}{\tau_2} \{\mu_2+\mathrm{logit} (x)\}\right].
\end{align*}

Accordingly, the HSAUC is defined by 
\begin{align*}
HSAUC(\boldsymbol{\mu}, \boldsymbol{\Omega}) 
= \int_{0}^{1}HSROC(x; \boldsymbol{\mu}, \boldsymbol{\Omega})dx.
\end{align*}

\end{appendix}

\clearpage

\beginsupplement
















\maketitle


\newpage

\section*{Supplementary Material for ``A likelihood-based sensitivity analysis for publication bias on summary ROC in meta-analysis of diagnostic test accuracy''}
\section*{1 Additional Application}

%

\begin{figure}[!htb]
\centering\includegraphics[width=0.8\columnwidth]{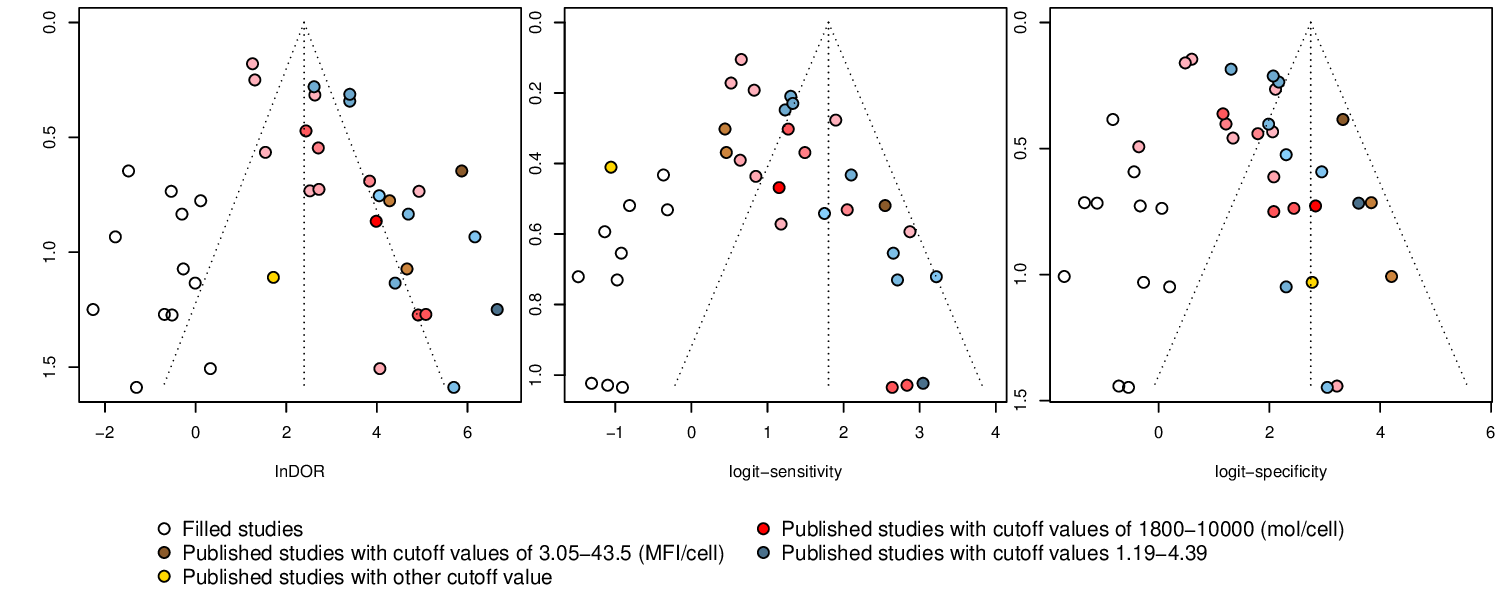}
\caption{
The trim-and-fill results on the lnDOR, logit-transformed sensitivity, and logit-transformed specificity in CD64 example.
The dashed lines are the estimates of the trim-and-fill adjustment.
The colors changing from light to dark indicate the increasing values.}
\label{fig:CD64-FUNNEL}
\end{figure}

In this section, 
we presented a second example to illustrate the proposed sensitivity analysis for publication bias using the HSROC curve or the HSAUC.
The meta-analysis of 27 studies quantitatively summarized the test accuracy of the neutrophil CD64 expression as a biomarker in differentiating bacterial infected patients from other non-infected patients 
and concluded that neutrophil CD64 expression could be a promising and meaningful biomarker for diagnosing bacterial infection with high sensitivity, specificity, and the SAUC.\cite{Li2013}
%
We draw the trim-and-fill methods on the lnDOR, logit-transformed sensitivity, and logit-transformed specificity.
The corresponding results in Figure \ref{fig:CD64-FUNNEL} suggested selective publication.
To investigate the impact of publication bias on the estimated HSROC curves or the HSAUC,
we considered the marginal probability of selective publication as $p=1,0.8,0.6,0.4$ 
under four selective publication mechanisms of $(\hat c_1, \hat c_2)$, $(c_1, c_2) = (1/\sqrt{2},1/\sqrt{2})$, $(1,0)$, and $(0,1)$.

\begin{figure}[!htb]
\centering\includegraphics[width=0.8\columnwidth]{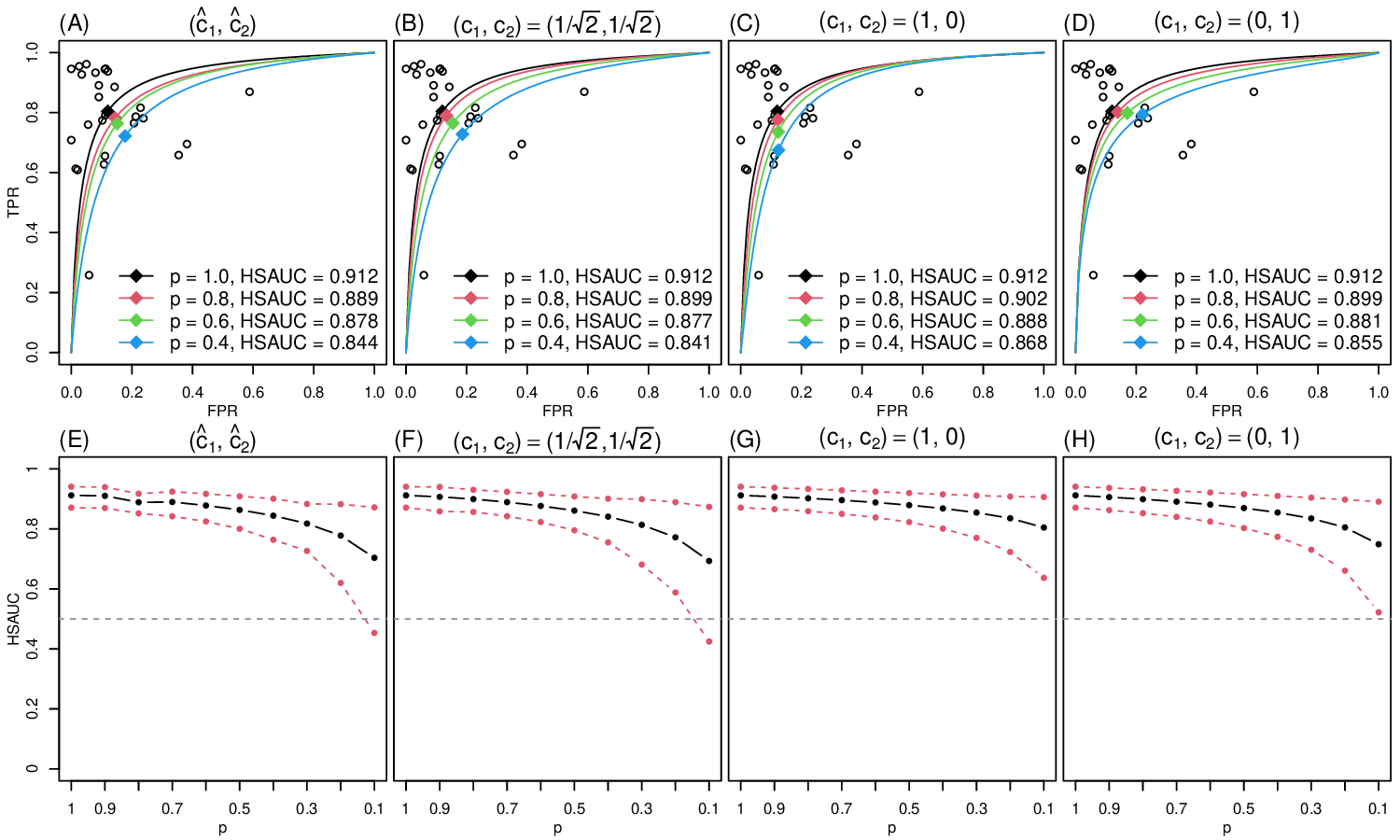}
\caption{
The estimated HSROC curves under four scenarios of selective publication mechanism in CD64 example.
The circle points are the observed FPR and TPR pairs from 27 primary studies.
The diamond points are the estimated SOP.
In panel (A), 
$(\hat{c}_1, \hat{c}_2) = (0.597, 0.802), (0.756, 0.655), (0.794, 0.608)$ given $p = 0.8, 0.6, 0.4$, respectively.}
\label{fig:CD64-SROC}
\end{figure}

\begin{figure}[!htb]
\centering\includegraphics[width=0.8\columnwidth]{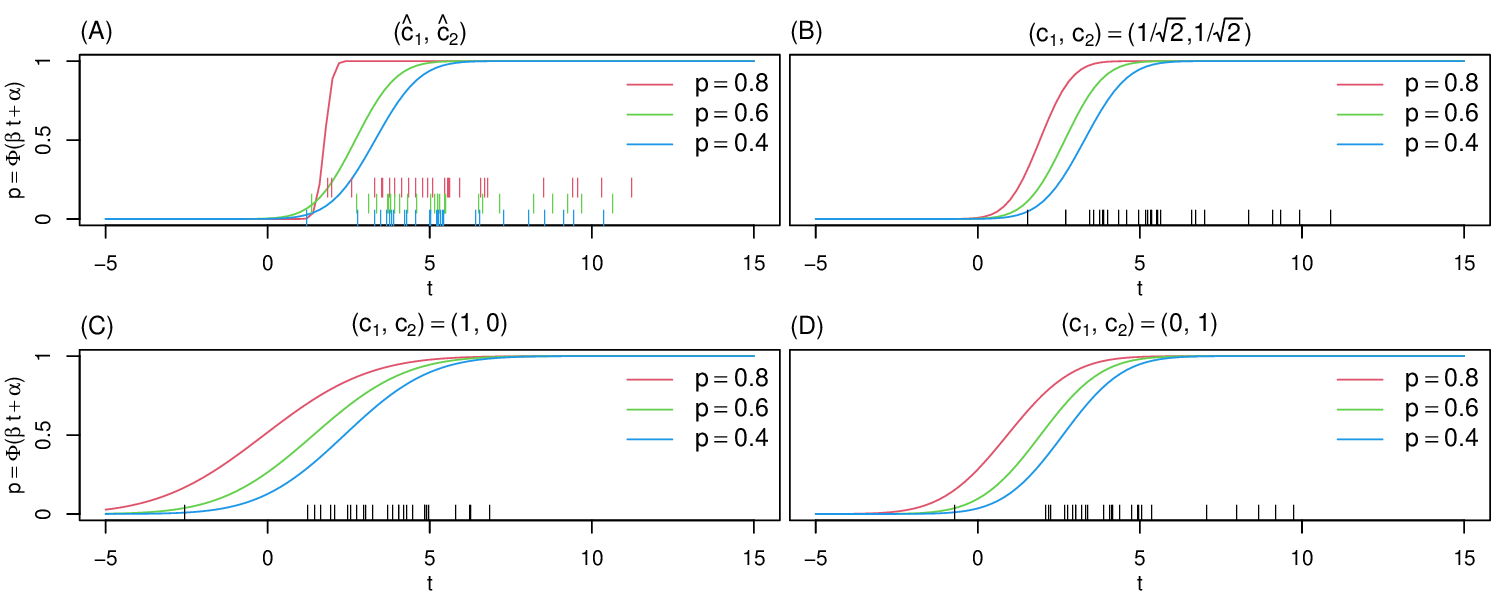}
\caption{
The estimated probit selection function $a(t_i)$ given different $p$'s in CD64 example.
The vertical lines at the bottom are the estimated $t$-type statistics from 27 primary studies.
}
\label{fig:CD64-PROBIT}
\end{figure}

Regarding $(c_1, c_2)$ as unknown,  
the HSAUC by the proposed method decreased from 0.912 to 0.844 as $p$ decreased,
as shown in panel (A) of Figure \ref{fig:CD64-SROC}.
When $p = 0.8$, differently from the example of IVD in Section 4 of the main paper, the relationships between $\hat c_1$ and $\hat c_2$ were dependent on $p$;
when $p=0.8$, $\hat c_1 < \hat c_2$, whereas when $p=0.4$, $\hat c_1 > \hat c_2$.
Despite of this inconsistency among $(\hat c_1,\hat c_2)$'s, 
the results with $(\hat c_1,\hat c_2)$ were similar to those with lnDOR-based selective publication, $(c_1, c_2) = (1/\sqrt{2}, 1/\sqrt{2})$.
When the significance of sensitivity or specificity was assumed to be the determinant of the selective publication mechanism,
the estimated HSAUC decreased from 0.912 to 0.868 or 0.855, respectively.
When $p$ decreased, the trajectories of the SOP went into the direction of potentially unpublished studies under different selective publication mechanisms.
The estimated TPR and FPR with 95\% confidence intervals are presented in Table \ref{tab:CD64}.
The estimates of the selection function are presented in Figure \ref{fig:CD64-PROBIT}.
%

%
As shown in Figure \ref{fig:CD64-SROC}, 
under four assumed selective publication mechanisms, 
the estimated HSAUC maintained statistically significant even if the marginal probability of selective publication was low as $p=0.2$, 
which corresponded to the presence of 108 unpublished studies.
To summary, the proposed sensitivity analysis strengthened the conclusion that neutrophil CD64 expression is a significant biomarker in diagnosing bacterial infection.

\section*{2 Simulation studies with \texorpdfstring{$(\tau_1^2, \tau_2^2) = (0.5, 0.5)$}{}}

As mentioned in the end of Section 5.1 of the main paper, the estimates of SAUC were consistent and the proposed methods removed biases regardless of the values of $(\tau_1^2, \tau_2^2)$. 
To confirm the influence of small $(\tau_1^2, \tau_2^2)$ on the simulation studies,
we generated additional six sets of simulation datasets with $(\tau_1^2, \tau_2^2) = (0.5,0.5)$.
The settings of the parameters are presented in Table \ref{tab:sceall}.
Values of $\mu_1, \mu_2,\rho, \beta$ were same as the original simulation scenarios in Table 3 of the main paper.
With $(\tau_1^2, \tau_2^2)$ modified, the values of SAUC were changed as given in Table \ref{tab:sceall}.
To make marginal probability $p$ approximate 0.7, $\alpha$'s were calculated correspondingly.

\clearpage





\begin{sidewaystable}

\caption{\label{tab:IVD}IVD example: the estimation under four selective publication mechanisms}
\centering
\begin{tabular}[!htb]{rrrrrrrrrrrr}
\toprule
$(c_1, \,c_2)$ & $p$ & SAUC (95\%CI) & Se (95\%CI) & Sp (95\%CI) & $\beta$ (95\%CI) & $\alpha_p$ & $\mu_1$ & $\mu_2$ & $\tau_1$ & $\tau_2$ & $\rho$\\
\midrule
$(\hat{c}_1, \hat{c}_2)$ & 1.0 & 0.874 (0.806, 0.920) & 0.815 (0.757, 0.862) & 0.861 (0.821, 0.893) &  &  & 1.484 & 1.824 & 0.597 & 0.826 & -0.424\\
 & 0.8 & 0.864 (0.790, 0.915) & 0.808 (0.751, 0.855) & 0.857 (0.816, 0.891) & 2.000 (-0.351, 4.351) & -3.451 & 1.440 & 1.794 & 0.601 & 0.833 & -0.388\\
 & 0.6 & 0.834 (0.740, 0.899) & 0.787 (0.724, 0.839) & 0.841 (0.794, 0.879) & 2.000 (-0.253, 4.253) & -4.857 & 1.309 & 1.665 & 0.590 & 0.859 & -0.312\\
 & 0.4 & 0.786 (0.648, 0.880) & 0.757 (0.676, 0.823) & 0.806 (0.739, 0.860) & 2.000 (0.528, 3.472) & -5.709 & 1.137 & 1.427 & 0.587 & 0.911 & -0.194\\
\addlinespace
$(0.7, 0.7)$ & 1.0 & 0.874 (0.806, 0.920) & 0.815 (0.757, 0.862) & 0.861 (0.821, 0.893) &  &  & 1.484 & 1.824 & 0.597 & 0.826 & -0.424\\
 & 0.8 & 0.864 (0.792, 0.915) & 0.809 (0.752, 0.855) & 0.857 (0.817, 0.890) & 2.000 (-0.132, 4.132) & -3.636 & 1.442 & 1.791 & 0.600 & 0.835 & -0.388\\
 & 0.6 & 0.834 (0.741, 0.897) & 0.786 (0.729, 0.834) & 0.842 (0.797, 0.878) & 2.000 (0.261, 3.739) & -4.757 & 1.301 & 1.671 & 0.592 & 0.857 & -0.310\\
 & 0.4 & 0.778 (0.636, 0.875) & 0.749 (0.676, 0.810) & 0.812 (0.750, 0.862) & 2.000 (0.525, 3.475) & -5.424 & 1.091 & 1.464 & 0.605 & 0.897 & -0.176\\
\addlinespace
$(1, 0)$ & 1.0 & 0.874 (0.806, 0.920) & 0.815 (0.757, 0.862) & 0.861 (0.821, 0.893) &  &  & 1.484 & 1.824 & 0.597 & 0.826 & -0.424\\
 & 0.8 & 0.859 (0.779, 0.913) & 0.786 (0.725, 0.836) & 0.866 (0.827, 0.898) & 1.149 (0.011, 2.288) & -0.204 & 1.302 & 1.869 & 0.677 & 0.833 & -0.413\\
 & 0.6 & 0.835 (0.737, 0.902) & 0.744 (0.669, 0.806) & 0.873 (0.831, 0.905) & 1.057 (0.317, 1.798) & -0.858 & 1.064 & 1.925 & 0.731 & 0.839 & -0.411\\
 & 0.4 & 0.800 (0.667, 0.889) & 0.682 (0.561, 0.782) & 0.880 (0.833, 0.915) & 1.002 (0.345, 1.658) & -1.315 & 0.762 & 1.992 & 0.784 & 0.844 & -0.408\\
\addlinespace
$(0, 1)$ & 1.0 & 0.874 (0.806, 0.920) & 0.815 (0.757, 0.862) & 0.861 (0.821, 0.893) &  &  & 1.484 & 1.824 & 0.597 & 0.826 & -0.424\\
 & 0.8 & 0.871 (0.803, 0.918) & 0.825 (0.766, 0.871) & 0.831 (0.780, 0.871) & 0.669 (-0.045, 1.383) & -1.214 & 1.548 & 1.589 & 0.614 & 0.973 & -0.467\\
 & 0.6 & 0.870 (0.802, 0.917) & 0.836 (0.774, 0.884) & 0.782 (0.704, 0.844) & 0.624 (0.081, 1.167) & -1.744 & 1.632 & 1.276 & 0.629 & 1.102 & -0.504\\
 & 0.4 & 0.869 (0.801, 0.916) & 0.852 (0.778, 0.904) & 0.697 (0.545, 0.816) & 0.590 (0.089, 1.090) & -2.062 & 1.749 & 0.834 & 0.648 & 1.250 & -0.544\\
\bottomrule
\end{tabular}
\end{sidewaystable}

\begin{sidewaystable}

\caption{\label{tab:CD64}CD64 example: the estimation under four selective publication mechanisms}
\centering
\begin{tabular}[!htb]{rrrrrrrrrrrr}
\toprule
$(c_1, \,c_2)$ & $p$ & HSAUC (95\%CI) & Se (95\%CI) & Sp (95\%CI) & $\beta$ (95\%CI) & $\alpha_p$ & $\mu_1$ & $\mu_2$ & $\tau_1$ & $\tau_2$ & $\rho$\\
\midrule
$(\hat{c}_1, \hat{c}_2)$ & 1.0 & 0.912 (0.871, 0.941) & 0.804 (0.742, 0.855) & 0.880 (0.831, 0.916) &  &  & 1.413 & 1.989 & 0.794 & 0.854 & 0.126\\
 & 0.8 & 0.889 (0.852, 0.917) & 0.782 (0.723, 0.832) & 0.855 (0.810, 0.891) & 5.000 (-16.845, 26.845) & -8.774 & 1.278 & 1.775 & 0.852 & 0.936 & 0.296\\
 & 0.6 & 0.878 (0.825, 0.917) & 0.764 (0.694, 0.822) & 0.848 (0.787, 0.894) & 1.004 (0.114, 1.893) & -2.713 & 1.173 & 1.721 & 0.858 & 0.894 & 0.288\\
 & 0.4 & 0.844 (0.764, 0.901) & 0.722 (0.623, 0.803) & 0.822 (0.740, 0.883) & 0.917 (0.332, 1.502) & -3.038 & 0.953 & 1.533 & 0.890 & 0.902 & 0.349\\
\addlinespace
$(0.7, 0.7)$ & 1.0 & 0.912 (0.871, 0.941) & 0.804 (0.742, 0.855) & 0.880 (0.831, 0.916) &  &  & 1.413 & 1.989 & 0.794 & 0.854 & 0.126\\
 & 0.8 & 0.899 (0.857, 0.931) & 0.790 (0.728, 0.841) & 0.867 (0.817, 0.905) & 1.272 (-0.185, 2.729) & -2.433 & 1.325 & 1.874 & 0.829 & 0.882 & 0.220\\
 & 0.6 & 0.877 (0.823, 0.916) & 0.765 (0.694, 0.823) & 0.845 (0.787, 0.890) & 1.059 (0.125, 1.993) & -2.843 & 1.179 & 1.700 & 0.855 & 0.904 & 0.296\\
 & 0.4 & 0.841 (0.755, 0.901) & 0.728 (0.632, 0.807) & 0.814 (0.730, 0.876) & 0.954 (0.290, 1.617) & -3.137 & 0.986 & 1.476 & 0.879 & 0.925 & 0.365\\
\addlinespace
$(1, 0)$ & 1.0 & 0.912 (0.871, 0.941) & 0.804 (0.742, 0.855) & 0.880 (0.831, 0.916) &  &  & 1.413 & 1.989 & 0.794 & 0.854 & 0.126\\
 & 0.8 & 0.902 (0.859, 0.933) & 0.776 (0.707, 0.833) & 0.878 (0.827, 0.915) & 0.392 (-0.006, 0.790) & 0.044 & 1.245 & 1.973 & 0.870 & 0.856 & 0.120\\
 & 0.6 & 0.888 (0.838, 0.924) & 0.736 (0.648, 0.808) & 0.876 (0.819, 0.917) & 0.447 (0.119, 0.775) & -0.630 & 1.025 & 1.958 & 0.928 & 0.858 & 0.106\\
 & 0.4 & 0.868 (0.801, 0.915) & 0.675 (0.543, 0.784) & 0.875 (0.804, 0.923) & 0.482 (0.192, 0.771) & -1.139 & 0.731 & 1.950 & 0.985 & 0.859 & 0.082\\
\addlinespace
$(0, 1)$ & 1.0 & 0.912 (0.871, 0.941) & 0.804 (0.742, 0.855) & 0.880 (0.831, 0.916) &  &  & 1.413 & 1.989 & 0.794 & 0.854 & 0.126\\
 & 0.8 & 0.899 (0.853, 0.932) & 0.802 (0.739, 0.852) & 0.860 (0.806, 0.901) & 0.631 (-0.003, 1.265) & -0.574 & 1.398 & 1.818 & 0.795 & 0.929 & 0.126\\
 & 0.6 & 0.881 (0.825, 0.921) & 0.799 (0.731, 0.853) & 0.829 (0.760, 0.882) & 0.675 (0.210, 1.139) & -1.303 & 1.378 & 1.582 & 0.795 & 0.995 & 0.131\\
 & 0.4 & 0.855 (0.774, 0.910) & 0.794 (0.712, 0.858) & 0.778 (0.659, 0.865) & 0.688 (0.255, 1.120) & -1.805 & 1.350 & 1.256 & 0.796 & 1.071 & 0.139\\
\bottomrule
\end{tabular}
\end{sidewaystable}

\begin{table}

\caption{\label{tab:sceall}Additional scenarios for simulation studies given $p \approx 0.7$}
\centering
\begin{tabular}[!hb]{rrrrrrrrrrrr}
\toprule
\multicolumn{1}{c}{ } & \multicolumn{1}{c}{ } & \multicolumn{1}{c}{ } & \multicolumn{1}{c}{ } & \multicolumn{1}{c}{ } & \multicolumn{1}{c}{ } & \multicolumn{1}{c}{ } & \multicolumn{1}{c}{ } & \multicolumn{1}{c}{ } & \multicolumn{1}{c}{$c_1 = c_2$} & \multicolumn{1}{c}{$c_1 = 1$} & \multicolumn{1}{c}{$c_1 = 0$} \\
\cmidrule(l{3pt}r{3pt}){10-10} \cmidrule(l{3pt}r{3pt}){11-11} \cmidrule(l{3pt}r{3pt}){12-12}
No. & SAUC & $\mu_1$ & $\mu_2$ & $\tau_1^2$ & $\tau_2^2$ & $\tau_{12}$ & $\rho$ & $\beta$ & $\alpha_{0.7}$ & $\alpha_{0.7}$ & $\alpha_{0.7}$\\
\midrule
7 & 0.620 & 0.000 & 1.735 & 0.5 & 0.5 & -0.15 & -0.3 & 0.5 & -0.423 & 0.794 & -0.993\\
8 & 0.702 & 0.000 & 1.735 & 0.5 & 0.5 & -0.30 & -0.6 & 0.5 & -0.461 & 0.795 & -0.996\\
9 & 0.846 & 1.386 & 1.386 & 0.5 & 0.5 & -0.15 & -0.3 & 0.5 & -1.003 & -0.698 & -0.697\\
10 & 0.864 & 1.386 & 1.386 & 0.5 & 0.5 & -0.30 & -0.6 & 0.5 & -1.032 & -0.701 & -0.698\\
11 & 0.877 & 2.197 & -0.405 & 0.5 & 0.5 & -0.15 & -0.3 & 0.5 & -0.457 & -1.362 & 1.342\\
12 & 0.835 & 2.197 & -0.405 & 0.5 & 0.5 & -0.30 & -0.6 & 0.5 & -0.492 & -1.362 & 1.335\\
\bottomrule
\end{tabular}
\end{table}

\begin{table}

\caption{\label{tab:c11}Summary of the SAUC estimates under the true selective publication mechanism of $(c_1, c_2) = (1/\sqrt{2}, 1/\sqrt{2})$}
\centering
\begin{tabular}[t]{rrrrrrr}
\toprule
\multicolumn{1}{c}{} & \multicolumn{1}{c}{} & \multicolumn{1}{c}{} & \multicolumn{1}{c}{$S = 15$} & \multicolumn{1}{c}{$S = 25$} & \multicolumn{1}{c}{$S = 50$} & \multicolumn{1}{c}{$S = 200$} \\
\cmidrule(l{3pt}r{3pt}){4-4} \cmidrule(l{3pt}r{3pt}){5-5} \cmidrule(l{3pt}r{3pt}){6-6} \cmidrule(l{3pt}r{3pt}){7-7}
No. &  Methods & True & Median (Q1, Q3) & Median (Q1, Q3) & Median (Q1, Q3) & Median (Q1, Q3)\\
\midrule
7 & Proposed $(\hat{c}_1, \hat{c}_2)$ & 62.0 & 66.1 (45.2, 77.2) & 64.3 (50.7, 74.1) & 62.5 (52.0, 70.4) & 61.8 (55.9, 65.9)\\
 & Proposed $(c_1 = c_2)$ &  & 65.8 (42.5, 76.6) & 64.4 (51.5, 74.3) & 63.3 (53.9, 70.6) & 62.2 (57.2, 66.0)\\
 & Proposed $(c_1 = 1)$ &  & 69.3 (51.1, 78.3) & 69.5 (59.8, 76.3) & 69.2 (63.5, 73.9) & 68.4 (65.7, 70.7)\\
 & Heckman-type &  & 67.5 (52.1, 76.7) & 68.4 (57.9, 74.4) & 66.8 (60.3, 71.9) & 66.5 (62.9, 69.3)\\
 & Reitsma$_O$ &  & 69.8 (53.6, 78.7) & 69.7 (61.1, 76.6) & 69.4 (64.1, 74.3) & 69.1 (66.3, 71.2)\\
 & Reitsma$_P$ &  & 63.3 (48.5, 73.0) & 62.2 (53.7, 69.6) & 62.0 (56.5, 67.1) & 61.8 (59.1, 64.6)\\
\addlinespace
8 & Proposed $(\hat{c}_1, \hat{c}_2)$ & 70.2 & 72.4 (61.2, 78.9) & 71.0 (62.5, 76.3) & 71.1 (65.3, 75.1) & 70.2 (67.7, 72.4)\\
 & Proposed $(c_1 = c_2)$ &  & 72.0 (59.7, 78.6) & 71.0 (62.7, 76.6) & 71.0 (65.3, 75.1) & 70.3 (67.9, 72.3)\\
 & Proposed $(c_1 = 1)$ &  & 74.1 (65.9, 79.8) & 73.4 (67.8, 77.9) & 73.6 (70.4, 76.7) & 73.4 (71.9, 75.2)\\
 & Heckman-type &  & 72.5 (63.6, 78.0) & 71.8 (64.6, 76.3) & 71.9 (67.2, 75.2) & 71.7 (69.6, 73.6)\\
 & Reitsma$_O$ &  & 74.4 (66.5, 79.9) & 73.7 (68.8, 78.1) & 74.1 (70.7, 77.0) & 73.8 (72.3, 75.2)\\
 & Reitsma$_P$ &  & 71.5 (63.5, 76.9) & 70.5 (64.7, 74.7) & 70.3 (66.8, 73.4) & 70.1 (68.5, 71.7)\\
\addlinespace
9 & Proposed $(\hat{c}_1, \hat{c}_2)$ & 84.6 & 84.1 (77.9, 87.5) & 84.6 (79.7, 87.2) & 84.6 (81.5, 86.6) & 84.8 (83.3, 85.9)\\
 & Proposed $(c_1 = c_2)$ &  & 83.8 (76.7, 87.3) & 84.6 (79.1, 87.1) & 84.7 (81.4, 86.7) & 84.7 (83.2, 85.8)\\
 & Proposed $(c_1 = 1)$ &  & 84.4 (78.6, 87.7) & 85.1 (81.0, 87.4) & 85.5 (83.3, 87.1) & 85.8 (84.6, 86.8)\\
 & Heckman-type &  & 84.8 (79.4, 87.9) & 85.5 (82.1, 87.5) & 86.1 (83.9, 87.5) & 86.2 (85.1, 87.0)\\
 & Reitsma$_O$ &  & 85.5 (80.3, 88.4) & 86.2 (83.2, 88.2) & 86.7 (84.8, 88.1) & 86.9 (85.9, 87.6)\\
 & Reitsma$_P$ &  & 83.6 (78.1, 86.4) & 84.2 (80.7, 86.5) & 84.5 (82.5, 86.0) & 84.6 (83.7, 85.5)\\
\addlinespace
10 & Proposed $(\hat{c}_1, \hat{c}_2)$ & 86.4 & 85.9 (82.8, 87.9) & 86.3 (84.2, 87.8) & 86.4 (85.1, 87.4) & 86.4 (85.9, 87.0)\\
 & Proposed $(c_1 = c_2)$ &  & 85.9 (82.3, 87.8) & 86.3 (84.2, 87.8) & 86.4 (85.1, 87.5) & 86.4 (85.9, 87.0)\\
 & Proposed $(c_1 = 1)$ &  & 86.2 (83.3, 88.0) & 86.5 (84.6, 88.0) & 86.8 (85.6, 87.7) & 86.9 (86.4, 87.5)\\
 & Heckman-type &  & 86.3 (83.2, 88.0) & 86.5 (84.8, 88.0) & 86.8 (85.6, 87.7) & 86.9 (86.4, 87.4)\\
 & Reitsma$_O$ &  & 87.0 (84.4, 88.6) & 87.3 (85.8, 88.5) & 87.6 (86.6, 88.4) & 87.7 (87.3, 88.1)\\
 & Reitsma$_P$ &  & 85.8 (83.6, 87.5) & 86.1 (84.5, 87.5) & 86.3 (85.2, 87.2) & 86.4 (85.9, 86.8)\\
\addlinespace
11 & Proposed $(\hat{c}_1, \hat{c}_2)$ & 87.7 & 85.7 (79.5, 89.4) & 86.7 (82.5, 89.4) & 87.6 (84.9, 89.4) & 87.8 (86.5, 88.8)\\
 & Proposed $(c_1 = c_2)$ &  & 85.4 (78.8, 89.3) & 86.5 (82.6, 89.4) & 87.4 (85.1, 89.3) & 87.7 (86.6, 88.7)\\
 & Proposed $(c_1 = 1)$ &  & 84.5 (77.2, 88.9) & 85.6 (80.8, 89.1) & 86.3 (83.5, 88.7) & 86.9 (85.5, 88.0)\\
 & Heckman-type &  & 86.0 (80.0, 89.6) & 86.6 (82.5, 89.6) & 87.2 (84.8, 89.5) & 87.6 (86.3, 88.7)\\
 & Reitsma$_O$ &  & 86.0 (79.1, 89.8) & 87.1 (83.0, 89.9) & 87.7 (85.4, 89.8) & 88.0 (87.0, 89.1)\\
 & Reitsma$_P$ &  & 86.1 (80.9, 89.2) & 87.2 (83.8, 89.3) & 87.5 (85.5, 89.1) & 87.7 (86.7, 88.5)\\
\addlinespace
12 & Proposed $(\hat{c}_1, \hat{c}_2)$ & 83.5 & 82.6 (76.0, 87.2) & 83.8 (79.0, 87.4) & 84.0 (81.1, 86.8) & 83.7 (82.2, 85.3)\\
 & Proposed $(c_1 = c_2)$ &  & 82.3 (76.0, 86.9) & 83.6 (79.3, 87.1) & 83.7 (80.8, 86.5) & 83.5 (82.1, 84.8)\\
 & Proposed $(c_1 = 1)$ &  & 81.0 (74.6, 86.3) & 82.2 (77.7, 86.1) & 82.7 (79.8, 85.5) & 82.8 (81.5, 84.2)\\
 & Heckman-type &  & 82.5 (76.7, 87.1) & 83.8 (79.5, 86.7) & 83.5 (80.6, 86.3) & 83.5 (82.2, 84.9)\\
 & Reitsma$_O$ &  & 82.5 (76.9, 87.1) & 83.9 (80.1, 87.1) & 84.1 (81.5, 86.6) & 84.2 (82.9, 85.4)\\
 & Reitsma$_P$ &  & 82.5 (76.8, 86.7) & 83.3 (79.6, 86.5) & 83.3 (80.8, 85.7) & 83.4 (82.2, 84.5)\\
\bottomrule
\end{tabular}
\begin{tablenotes}
\item 
Median with 25th empirical percentile (Q1) and 75th empirical percentile (Q3) and are reported. 
No. corresponds to the scenario    number.
$S$ denotes the number of the population studies. 
True denotes the the true value of the SAUC.
Proposed $(\hat c_1, \hat c_2)$, Proposed $(c_1 = c_2)$, and Proposed $(c_1 = 1)$ denote 
the proposed method that estimates $(c_1, c_2)$, 
correctly specifies $(c_1, c_2) = (1/\sqrt{2}, 1/\sqrt{2})$, 
and misspecifies $(c_1, c_2) = (1,0)$, respectively;
Heckman-type denotes the method of Piao et al.;
Reitsma$_O$ and Reitsma$_P$ denote the Reitsma model 
based on $N$ published studies and $S$ population studies, respectively. 
All the entries are multiplied by 100.
\end{tablenotes}
\end{table}
 
\begin{table}

\caption{\label{tab:c10}Summary of the SAUC estimates under the true selective publication mechanism of $(c_1, c_2) = (1,0)$}
\centering
\begin{tabular}[t]{rrrrrrr}
\toprule
\multicolumn{1}{c}{} & \multicolumn{1}{c}{} & \multicolumn{1}{c}{} & \multicolumn{1}{c}{$S = 15$} & \multicolumn{1}{c}{$S = 25$} & \multicolumn{1}{c}{$S = 50$} & \multicolumn{1}{c}{$S = 200$} \\
\cmidrule(l{3pt}r{3pt}){4-4} \cmidrule(l{3pt}r{3pt}){5-5} \cmidrule(l{3pt}r{3pt}){6-6} \cmidrule(l{3pt}r{3pt}){7-7}
No. &   & True & Median (Q1, Q3) & Median (Q1, Q3) & Median (Q1, Q3) & Median (Q1, Q3)\\
\midrule
7 & Proposed $(\hat{c}_1, \hat{c}_2)$ & 62.0 & 63.2 (47.5, 74.7) & 63.1 (50.9, 71.5) & 62.9 (54.5, 69.2) & 63.7 (59.7, 67.0)\\
 & Proposed $(c_1 = 1)$ &  & 64.2 (47.1, 75.2) & 64.8 (51.3, 73.1) & 62.7 (54.9, 69.9) & 62.3 (57.9, 65.5)\\
 & Proposed $(c_1 = c_2)$ &  & 62.2 (48.0, 74.5) & 62.8 (51.3, 72.0) & 63.9 (56.4, 70.1) & 65.2 (62.1, 68.0)\\
 & Heckman-type &  & 61.3 (48.4, 71.9) & 62.4 (51.7, 69.9) & 60.8 (53.8, 66.8) & 59.9 (55.4, 63.4)\\
 & Reitsma$_O$ &  & 66.3 (51.9, 75.6) & 66.6 (56.4, 73.9) & 66.1 (60.8, 71.4) & 66.0 (63.0, 68.6)\\
 & Reitsma$_P$ &  & 63.3 (48.5, 73.0) & 62.2 (53.7, 69.6) & 62.0 (56.5, 67.1) & 61.8 (59.1, 64.6)\\
\addlinespace
8 & Proposed $(\hat{c}_1, \hat{c}_2)$ & 70.2 & 69.4 (57.5, 77.1) & 69.9 (60.4, 75.3) & 70.1 (64.3, 74.0) & 70.5 (68.4, 72.3)\\
 & Proposed $(c_1 = 1)$ &  & 71.3 (59.0, 77.9) & 71.1 (63.0, 75.9) & 71.1 (65.9, 74.8) & 70.3 (68.3, 72.1)\\
 & Proposed $(c_1 = c_2)$ &  & 69.3 (57.3, 77.1) & 69.5 (60.3, 75.0) & 70.4 (64.7, 74.5) & 71.2 (69.1, 72.9)\\
 & Heckman-type &  & 67.3 (55.9, 74.8) & 66.9 (58.0, 72.8) & 67.2 (59.7, 71.7) & 65.9 (61.7, 69.1)\\
 & Reitsma$_O$ &  & 71.6 (61.6, 78.0) & 71.7 (65.5, 76.0) & 71.8 (67.7, 75.2) & 71.7 (70.0, 73.3)\\
 & Reitsma$_P$ &  & 71.5 (63.5, 76.9) & 70.5 (64.7, 74.7) & 70.3 (66.8, 73.4) & 70.1 (68.5, 71.7)\\
\addlinespace
9 & Proposed $(\hat{c}_1, \hat{c}_2)$ & 84.6 & 82.9 (75.0, 86.8) & 84.1 (78.9, 86.7) & 84.2 (80.6, 86.4) & 84.5 (83.0, 85.5)\\
 & Proposed $(c_1 = 1)$ &  & 83.0 (75.4, 86.9) & 84.4 (80.2, 86.9) & 84.5 (81.2, 86.5) & 84.6 (83.4, 85.7)\\
 & Proposed $(c_1 = c_2)$ &  & 82.9 (74.1, 87.0) & 84.1 (79.2, 86.8) & 84.5 (81.0, 86.6) & 84.9 (83.5, 86.0)\\
 & Heckman-type &  & 83.3 (77.2, 87.2) & 84.8 (80.9, 87.0) & 85.1 (82.3, 86.8) & 85.3 (84.1, 86.3)\\
 & Reitsma$_O$ &  & 84.3 (77.8, 87.7) & 85.8 (82.2, 87.7) & 86.0 (83.6, 87.7) & 86.3 (85.3, 87.1)\\
 & Reitsma$_P$ &  & 83.6 (78.1, 86.4) & 84.2 (80.7, 86.5) & 84.5 (82.5, 86.0) & 84.6 (83.7, 85.5)\\
\addlinespace
10 & Proposed $(\hat{c}_1, \hat{c}_2)$ & 86.4 & 85.5 (81.9, 87.7) & 85.9 (83.7, 87.6) & 86.2 (84.9, 87.4) & 86.4 (85.8, 86.9)\\
 & Proposed $(c_1 = 1)$ &  & 85.5 (82.2, 87.8) & 86.1 (83.9, 87.7) & 86.4 (85.0, 87.4) & 86.4 (85.9, 87.0)\\
 & Proposed $(c_1 = c_2)$ &  & 85.5 (81.7, 87.6) & 86.1 (83.8, 87.8) & 86.5 (85.2, 87.7) & 86.7 (86.1, 87.3)\\
 & Heckman-type &  & 85.5 (81.6, 87.5) & 85.9 (83.7, 87.6) & 86.4 (84.8, 87.4) & 86.4 (85.6, 87.0)\\
 & Reitsma$_O$ &  & 86.4 (83.6, 88.3) & 87.0 (85.1, 88.4) & 87.4 (86.4, 88.3) & 87.6 (87.1, 88.0)\\
 & Reitsma$_P$ &  & 85.8 (83.6, 87.5) & 86.1 (84.5, 87.5) & 86.3 (85.2, 87.2) & 86.4 (85.9, 86.8)\\
\addlinespace
11 & Proposed $(\hat{c}_1, \hat{c}_2)$ & 87.7 & 86.7 (80.0, 90.1) & 88.0 (84.2, 90.3) & 88.6 (86.3, 90.0) & 88.2 (86.9, 89.3)\\
 & Proposed $(c_1 = 1)$ &  & 85.5 (77.8, 89.3) & 86.8 (82.3, 89.7) & 87.8 (85.1, 89.5) & 87.7 (86.5, 88.8)\\
 & Proposed $(c_1 = c_2)$ &  & 86.7 (79.8, 90.0) & 88.4 (85.0, 90.5) & 89.0 (87.1, 90.3) & 89.2 (88.3, 89.8)\\
 & Heckman-type &  & 87.4 (81.5, 90.6) & 88.8 (85.1, 90.9) & 89.3 (87.0, 90.6) & 89.3 (88.3, 90.2)\\
 & Reitsma$_O$ &  & 86.9 (79.4, 90.3) & 88.5 (84.7, 90.8) & 89.2 (86.9, 90.7) & 89.3 (88.4, 90.1)\\
 & Reitsma$_P$ &  & 86.1 (80.9, 89.2) & 87.2 (83.8, 89.3) & 87.5 (85.5, 89.1) & 87.7 (86.7, 88.5)\\
\addlinespace
12 & Proposed $(\hat{c}_1, \hat{c}_2)$ & 83.5 & 83.2 (75.7, 88.2) & 84.9 (79.4, 88.4) & 84.7 (81.2, 87.6) & 84.0 (82.2, 85.6)\\
 & Proposed $(c_1 = 1)$ &  & 81.7 (74.6, 86.9) & 83.6 (78.4, 87.3) & 83.7 (80.5, 86.3) & 83.5 (82.0, 84.8)\\
 & Proposed $(c_1 = c_2)$ &  & 83.7 (76.5, 88.3) & 85.1 (80.5, 88.4) & 85.5 (82.4, 87.9) & 85.2 (83.8, 86.4)\\
 & Heckman-type &  & 84.5 (78.0, 88.8) & 85.3 (80.5, 88.6) & 85.8 (82.8, 88.3) & 85.7 (84.3, 87.1)\\
 & Reitsma$_O$ &  & 83.0 (76.2, 88.1) & 84.9 (80.3, 88.2) & 85.4 (82.3, 87.7) & 85.3 (84.0, 86.5)\\
 & Reitsma$_P$ &  & 82.5 (76.8, 86.7) & 83.3 (79.6, 86.5) & 83.3 (80.8, 85.7) & 83.4 (82.2, 84.5)\\
\bottomrule
\end{tabular}
\begin{tablenotes}
\item 
Median with 25th empirical percentile (Q1) and 75th empirical percentile (Q3) are reported. 
No. corresponds to the scenario    number.
$S$ denotes the number of population studies.
True denotes the true value of the SAUC.
Proposed $(\hat c_1, \hat c_2)$, Proposed $(c_1 = 1)$, and Proposed $(c_1 = c_2)$ denote 
the proposed method that estimates $(c_1, c_2)$, 
correctly specifies $(c_1, c_2)=(1,0)$, 
and misspecifies $(c_1, c_2)=(1/\sqrt{2}, 1/\sqrt{2})$, respectively;
Heckman-type denotes the method of Piao et al.;
Reitsma$_O$ and Reitsma$_P$ denote the Reitsma model 
based on $N$ published studies and $S$ population studies, respectively. 
All the entries are multiplied by 100.
\end{tablenotes}
\end{table}
 
\begin{table}

\caption{\label{tab:c01}Summary of the SAUC estimates under the true selective publication mechanism of $(c_1, c_2) = (0,1)$}
\centering
\begin{tabular}[t]{rrrrrrr}
\toprule
\multicolumn{1}{c}{} & \multicolumn{1}{c}{} & \multicolumn{1}{c}{} & \multicolumn{1}{c}{$S = 15$} & \multicolumn{1}{c}{$S = 25$} & \multicolumn{1}{c}{$S = 50$} & \multicolumn{1}{c}{$S = 200$} \\
\cmidrule(l{3pt}r{3pt}){4-4} \cmidrule(l{3pt}r{3pt}){5-5} \cmidrule(l{3pt}r{3pt}){6-6} \cmidrule(l{3pt}r{3pt}){7-7}
No. &   & True & Median (Q1, Q3) & Median (Q1, Q3) & Median (Q1, Q3) & Median (Q1, Q3)\\
\midrule
7 & Proposed $(\hat{c}_1, \hat{c}_2)$ & 62.0 & 56.3 (33.3, 72.5) & 56.0 (39.8, 68.9) & 55.9 (42.6, 65.3) & 58.2 (52.8, 63.2)\\
 & Proposed $(c_1 = 0)$ &  & 62.2 (41.4, 74.4) & 62.7 (51.5, 72.3) & 62.4 (54.7, 68.7) & 61.9 (57.8, 65.2)\\
 & Proposed $(c_1 = c_2)$ &  & 54.8 (32.1, 72.3) & 54.3 (38.3, 68.6) & 55.4 (42.2, 64.1) & 54.7 (49.1, 60.1)\\
 & Heckman-type &  & 59.5 (39.4, 73.0) & 60.2 (46.0, 70.6) & 60.1 (49.7, 67.4) & 59.2 (54.5, 63.5)\\
 & Reitsma$_O$ &  & 62.7 (40.8, 75.2) & 63.6 (51.2, 73.1) & 62.7 (55.2, 69.3) & 62.3 (58.3, 65.8)\\
 & Reitsma$_P$ &  & 63.3 (48.5, 73.0) & 62.2 (53.7, 69.6) & 62.0 (56.5, 67.1) & 61.8 (59.1, 64.6)\\
\addlinespace
8 & Proposed $(\hat{c}_1, \hat{c}_2)$ & 70.2 & 69.6 (51.8, 78.2) & 67.3 (53.8, 75.1) & 68.0 (60.3, 73.2) & 68.7 (65.8, 71.1)\\
 & Proposed $(c_1 = 0)$ &  & 71.2 (59.2, 78.9) & 69.9 (61.2, 75.8) & 70.6 (65.8, 74.4) & 70.1 (68.1, 72.1)\\
 & Proposed $(c_1 = c_2)$ &  & 69.1 (48.7, 78.0) & 66.7 (51.2, 75.2) & 67.9 (60.1, 73.2) & 67.1 (63.9, 70.1)\\
 & Heckman-type &  & 69.9 (55.7, 77.7) & 68.0 (57.8, 74.4) & 68.7 (62.3, 73.1) & 68.6 (65.3, 71.0)\\
 & Reitsma$_O$ &  & 72.2 (59.3, 79.3) & 70.6 (61.5, 76.5) & 71.3 (66.6, 75.1) & 70.8 (68.8, 72.8)\\
 & Reitsma$_P$ &  & 71.5 (63.5, 76.9) & 70.5 (64.7, 74.7) & 70.3 (66.8, 73.4) & 70.1 (68.5, 71.7)\\
\addlinespace
9 & Proposed $(\hat{c}_1, \hat{c}_2)$ & 84.6 & 82.1 (69.4, 86.6) & 82.9 (74.6, 86.6) & 83.5 (78.6, 86.0) & 84.1 (82.3, 85.4)\\
 & Proposed $(c_1 = 0)$ &  & 83.3 (74.0, 87.2) & 84.0 (78.3, 87.1) & 84.6 (81.4, 86.7) & 84.6 (83.3, 85.7)\\
 & Proposed $(c_1 = c_2)$ &  & 81.6 (68.0, 86.4) & 82.5 (74.1, 86.5) & 82.9 (77.9, 85.8) & 83.1 (80.8, 84.8)\\
 & Heckman-type &  & 82.2 (71.2, 86.4) & 82.9 (75.3, 86.2) & 83.3 (78.8, 85.9) & 83.2 (81.1, 84.8)\\
 & Reitsma$_O$ &  & 83.5 (73.8, 87.6) & 84.5 (78.3, 87.4) & 85.0 (81.6, 87.0) & 84.9 (83.6, 86.1)\\
 & Reitsma$_P$ &  & 83.6 (78.1, 86.4) & 84.2 (80.7, 86.5) & 84.5 (82.5, 86.0) & 84.6 (83.7, 85.5)\\
\addlinespace
10 & Proposed $(\hat{c}_1, \hat{c}_2)$ & 86.4 & 85.0 (80.6, 87.5) & 85.5 (82.5, 87.3) & 85.7 (83.9, 87.1) & 86.1 (85.4, 86.8)\\
 & Proposed $(c_1 = 0)$ &  & 85.5 (82.0, 87.9) & 86.1 (83.7, 87.7) & 86.3 (84.8, 87.4) & 86.4 (85.8, 87.0)\\
 & Proposed $(c_1 = c_2)$ &  & 85.0 (79.6, 87.5) & 85.3 (82.2, 87.3) & 85.5 (83.5, 87.1) & 85.8 (84.9, 86.5)\\
 & Heckman-type &  & 85.0 (80.1, 87.1) & 85.3 (81.7, 87.1) & 85.4 (83.1, 86.8) & 85.4 (84.5, 86.2)\\
 & Reitsma$_O$ &  & 85.9 (82.3, 88.1) & 86.4 (84.0, 88.0) & 86.7 (85.2, 87.8) & 86.8 (86.1, 87.3)\\
 & Reitsma$_P$ &  & 85.8 (83.6, 87.5) & 86.1 (84.5, 87.5) & 86.3 (85.2, 87.2) & 86.4 (85.9, 86.8)\\
\addlinespace
11 & Proposed $(\hat{c}_1, \hat{c}_2)$ & 87.7 & 82.4 (72.6, 87.4) & 83.8 (76.9, 87.7) & 86.1 (82.4, 88.3) & 87.1 (85.5, 88.3)\\
 & Proposed $(c_1 = 0)$ &  & 84.3 (76.4, 88.4) & 86.0 (80.1, 88.7) & 87.2 (84.5, 88.9) & 87.6 (86.4, 88.6)\\
 & Proposed $(c_1 = c_2)$ &  & 82.3 (72.6, 87.2) & 84.3 (77.6, 87.6) & 86.3 (82.9, 88.3) & 87.5 (86.1, 88.5)\\
 & Heckman-type &  & 83.5 (75.7, 87.9) & 85.1 (80.3, 88.1) & 86.2 (83.6, 88.1) & 87.1 (85.9, 88.1)\\
 & Reitsma$_O$ &  & 83.2 (73.1, 87.7) & 84.8 (77.0, 88.2) & 86.7 (83.2, 88.7) & 87.6 (86.2, 88.6)\\
 & Reitsma$_P$ &  & 86.1 (80.9, 89.2) & 87.2 (83.8, 89.3) & 87.5 (85.5, 89.1) & 87.7 (86.7, 88.5)\\
\addlinespace
12 & Proposed $(\hat{c}_1, \hat{c}_2)$ & 83.5 & 78.7 (69.4, 84.9) & 80.9 (75.1, 85.4) & 82.2 (78.3, 85.4) & 82.4 (80.7, 84.1)\\
 & Proposed $(c_1 = 0)$ &  & 80.9 (73.1, 85.9) & 82.8 (77.7, 86.3) & 83.4 (80.3, 86.0) & 83.4 (82.0, 84.8)\\
 & Proposed $(c_1 = c_2)$ &  & 78.7 (69.2, 84.9) & 81.1 (75.3, 85.3) & 82.6 (78.8, 85.4) & 82.8 (81.2, 84.3)\\
 & Heckman-type &  & 80.0 (72.4, 85.4) & 82.0 (76.9, 85.9) & 83.0 (79.7, 85.7) & 83.5 (81.6, 85.3)\\
 & Reitsma$_O$ &  & 79.0 (70.1, 85.1) & 81.4 (75.5, 85.7) & 82.8 (79.2, 85.6) & 82.8 (81.4, 84.3)\\
 & Reitsma$_P$ &  & 82.5 (76.8, 86.7) & 83.3 (79.6, 86.5) & 83.3 (80.8, 85.7) & 83.4 (82.2, 84.5)\\
\bottomrule
\end{tabular}
\begin{tablenotes}
\item 
Median with 25th empirical percentile (Q1) and 75th empirical percentile (Q3)) are reported. 
No. corresponds to the scenario number.
$S$ denotes the number of the population studies. 
True denotes the the true value of the SAUC.
Proposed $(\hat c_1, \hat c_2)$, Proposed $(c_1 = 1)$, and Proposed $(c_1 = c_2)$ denote 
the proposed method that estimates $(c_1, c_2)$, 
correctly specifies $(c_1, c_2) = (0,1)$, 
and misspecifies $(c_1, c_2) = (1/\sqrt{2}, 1/\sqrt{2})$, respectively;
Heckman-type denotes the method of Piao et al.;
Reitsma$_O$ and Reitsma$_P$ denote the Reitsma model 
based on $N$ published studies and $S$ population studies, respectively. 
All the entries are multiplied by 100.
\end{tablenotes}
\end{table}

\end{document}